\documentclass[aps,prr,superscriptaddress,twocolumn, showpacs]{revtex4}
\usepackage{dcolumn}
\usepackage{graphicx} 
\usepackage{amsmath}
\usepackage{bm}
\usepackage{xcolor}
\usepackage[T1]{fontenc}
\usepackage[utf8]{inputenc}
\usepackage{enumerate}

\begin{document}

\title{Worldwide bilateral geopolitical interactions network inferred from national disciplinary profiles}
\author{M.G.~Izzo \footnote{mizzo@sissa.it; izzo@diag.uniroma1.it}}
\affiliation{Universit\'a degli studi di Roma ''La Sapienza'', Dipartimento di Ingegneria Informatica Automatica e Gestionale Antonio Ruberti, Via Ariosto, 00185 Roma, Italy \\}
\affiliation{Istituto Italiano di Tecnologia - Center for Life Nanoscience, Viale Regina Elena, 291
00161 Roma, Italy \\}
\affiliation{Scuola Internazionale Superiore di Studi Avanzati, via Bonomea, 265 34136 Trieste, Italy \\}
\author{C.~Daraio}
\affiliation{Universit\'a degli studi di Roma ''La Sapienza'', Dipartimento di Ingegneria Informatica Automatica e Gestionale Antonio Ruberti, Via Ariosto, 00185 Roma, Italy \\}
\author{L.~Leuzzi}
\affiliation{Universit\'a degli studi di Roma ''La Sapienza'', Dipartimento di Fisica, Piazzale Aldo Moro 5, 00185 Roma, Italy \\}
\affiliation{Soft and Living Matter Lab, Institute of Nanotechnology, CNR-NANOTEC, Rome, Italy  \\}
\author{G.~Quaglia}
\affiliation{Universit\'a degli studi di Roma ''La Sapienza'', Dipartimento di Ingegneria Informatica Automatica e Gestionale Antonio Ruberti, Via Ariosto, 00185 Roma, Italy \\}
\author{G.~Ruocco}
\affiliation{Universit\'a degli studi di Roma ''La Sapienza'', Dipartimento di Fisica, Piazzale Aldo Moro 5, 00185 Roma, Italy \\}
\affiliation{Istituto Italiano di Tecnologia - Center for Life Nanoscience, Viale Regina Elena, 291
00161 Roma, Italy \\}

\date{\today}

\begin{abstract}
A disciplinary profile of a country is defined as the versor whose components are the numbers of articles produced in a given discipline divided the overall production of the country. Starting from the Essential Science Indicators (ESI) schema of classification of subject area, we obtained the yearly disciplinary profiles of a worldwide graph, where on each node sits a country, in the two time intervals $[1980-1988]$ and $[1992-2017]$, the fall of the Berlin Wall being the watershed. We analyse the empirical pairwise cross-correlation matrices of the time series of disciplinary profiles. The contrast with random matrix theory proves that, beyond measurement noise, the empirical cross-correlation matrices bring genuine information.  Arising from the Shannon theorem as the least-structured model consistent with the measured pairwise correlations, the stationary probability distribution of disciplinary profiles can be described by a Boltzmann distribution related to a generalized $n_d$-dimensional Heisenberg model. The set of network interactions of the Heisenberg model have been inferred and to it they have been applied two clusterization methods, hierarchical clustering and principal component analysis. On a geopolitical plane this allow to obtain a characterization of the worldwide bilateral interactions based on physical modeling. A simple geopolitical analysis reveals the consistency of the results obtained and gives a boost to deeper historical analysis. In order to obtain the optimal set of pairwise interactions we used a Pseudo-Likelihood approach. We analytically computed the Pseudo-Likelihood and its gradient. The analytical computations deserve interest in whatever inference Bayesian problem involving a $n_d$-dimensional Heisenberg model. 

\end{abstract}

\maketitle
\section{Introduction} \label{introduction}
The key role of high-level specialized knowledge in the nowadays international political relations is an almost undoubted shared opinion. Beyond common sense this idea has been formalized in a socio-political context by concepts such as \lq soft power' [\onlinecite{Nye}] or \lq ecomonic war' [\onlinecite{Esambert}]. Scientific research and technological innovation contribute to the production potential of a State, in turn promoting its economic and, consequently, political power. Losing the production potential can drain it towards an actual political dependence in the worldwide political scenario. The Lisbon European Council held in March 2000 aimed to launch the European Union towards a so-called knowledge-based economy [\onlinecite{Lisbon}], emphasizing how in the contemporary world the culture is somehow thought to be linked to the economy. The strong increase of scientific literature production of emergent countries, such as China or other South-East Asian countries, can be also read under this light. Even if part of the scientific research is actually financed by private companies and most of filed patents belong to them, the role of the State in this dynamics remains preponderant. It retains indeed power in terms of both substantial funding of cutting edge fields and long-term planning of research strategies. The provision of a regulatory environment aimed to support and protect research and innovation, e.g. patent policy, is a further prerogative of the State. Scientific disciplines but not only come into play in these geopolitical dynamics. The legislative system of a country can directly act as an attractiveness of foreign investments, generating national employment and ensuring a tax benefit. The cultural background cast by arts or humanities disciplines contributes to the territorial attractiveness of the country. Economic or computer science knowledge can furnish the background for the so-called economic intelligence [\onlinecite{Juillet}]. Big data analytics, which requires specific informatic, technological and scientific competences, can be crucial in the control and creation of information channels. 
National disciplinary profiles, intended as the high-level and shared knowledge acquired in a given range of specialized fields by a given country, and international relations are thus interconnected since high-level knowledge acquired by the country in a given field can be linked and can promote a specific international political power. The specialized knowledge, furthermore, supports the flexibility of the country in order to adapt to external changes or to promote them at the expense of less versatile countries. The link between national culture and international relations has been recently the focus of several historical analysis and specific events have been analyzed on this perspective [\onlinecite{Stelowska, Huseynov, Cruz}]. The support of such a socio-political hypothesis by quantitative methods has, however, not be achieved. Even the formalization of the matter in a quantitative framework by the definition of proper indices has been questioned [\onlinecite{Stelowska}]. Without going into the merits of the causes, consequences and validity of the political dynamics sketched above, we aim at establishing if this \textit{status quo} can be inferred from a quantitative analysis of the country-level creation of high-level specialized articles.
Though evolutionary processes of cultures and cultural traits have been studied in the past [\onlinecite{Feldman}], a quantitative and well assessed analysis of their relationship with worldwide geopolitical interactions is lacking. On the other hand, nowadays databases collecting a large amount of specialized manuscripts in different disciplines are available and this permits to obtain measurements of well-defined observables, e.g. bibliometric indicators, in a given interval of time. Since the bibliometric  indicators are the \textit{a priori} defined observables, grounding on which a modeling for social analysis can be built, particular care needs to be reserved to their definition and use in the appropriate context [\onlinecite{Ruocco}]. 

On the basis of the previous observations we assume that: (i) it is possible to define a disciplinary profile related to a given country as a vector of $n_d$ elements, being $n_d$ the total number of disciplines taken into account. Each vector component is the relative number of articles related to the corresponding discipline with respect to the total country's production; (ii) in the perspective of analyzing the role of specialized knowledge on the worldwide geopolitical interactions it makes sense to define the disciplinary profile at the country-level. These two points permit to identify the national disciplinary profiles as our observables with respect to the matter under exam. Once defined the observables, we quantify their correlations by analyzing their cross-correlation matrices. The comparison with Random Correlation Matrices (RMCs) properties and Random Matrix Theory (RMT) results [\onlinecite{Metha}] permits to assess that empirical cross-correlations contain genuine information, related to the characteristics of the underlying network of interactions leading the observed national disciplinary profiles. 
We further analyse the stationary in time of the genuine information. We then apply maximum-entropy methods derived from the Shannon theorem to model the maximum entropy probability distribution consistent with the measured cross-correlations, without further assumption on not-analyzed higher-order interactions terms. The resulting probability distribution is the Boltzmann distribution related to the Hamiltonian of a generalized Heisenberg model with $n_d$-dimensional spin variables. 
 Development of appropriate tools in the framework of the so-called Pseudo-Likelihood approach, adapted to the generalized Heisenberg model of $n_d$-dimensional spin variables, permitted us to infer the set of pairwise interactions between different countries. In order to assess the consistency of the inferred interactions and to apply to them clusterization methods, their properties have been contrasted with the ones of RMCs [\onlinecite{Metha,RandomMatrix1,RandomMatrix2}]. Finally, we apply to the inferred interactions set algorithms usually exploited in the analysis of cross-correlations matrices, Hierarchical Clustering (HC) and Principal Component Analysis (PCA) [\onlinecite{Barber1,Laio}]. We obtain a division in clusters of different countries based on their interactions profile with all the other countries. 

We choose to analyse separately two different intervals of time, i.e. $[1980-1988]$ and $[1992-2017]$, the fall of the Berlin Wall being the watershed between the two time intervals. The end of the Cold War marked a rebuilding of the geographic borders. To separate the analysis in the two time intervals is thus unavoidable. A comparison between the interaction profiles inferred in the two time intervals is, however, interesting because the end of the Cold War also marked a change on the nature of international relations. From a geopolitical and military, though not directly acted, plane they moved towards a more distinctly geoeconomic plane. The detailed preliminary analysis of the cross-correlations functions have been performed only for the data referring to the time-interval $[1992-2017]$.
For sake of simplicity we analyze only lower order correlations functions, i.e. pairwise. This choice could be \textit{a posteriori} verified if the number of acquisitions would be large enough to obtain a reliable distribution of the N-correlated variables, where N is the total number of variables, as specified in the text. This is not achieved in the present case. 
The preference of the bilateral interactions with respect to multilateral ones, however, in a geopolitical plane can be identified with a specific tendency of the contemporary world, in particular arising after the end of the Cold War, when the political and economic links of solidarity inside each blocs had not met anymore. The bilateralism of the international relations has been analyzed also in the context of soft power [\onlinecite{Becsi, Garten}]. 
Once consistency of the model and of the inference method has been verified in the statistical groundwork, it is performed a simple socio-political analysis. The inferred interactions reflect quite well the international relations drawn by grounding on only geopolitic and geoeconomics arguments.

The paper is organised as follows. Sec. \ref{Databases} contains a definition of the countries disciplinary profiles and describes the database from which the bibliometric data have been extracted and the classification schemes. In Sec. \ref{CC} it is described the analysis of empirical cross correlation matrices. In Sec. \ref{method} it is introduced the inference method, the probability density distribution of the countries disciplinary profiles and the related generalized Heisenberg model with multidimensional spin variables. Furthermore, the analytical computation of Pseudo-Likelihood and its gradient for the generalized Hesenberg model are presented. In Sec. \ref{network} the inferred interactions network is characterized applying to it HC and PCA. Sec. \ref{discusion} contains a simple geopolitical analysis of the inferred interactions set. Concluding remarks and outlooks are presented in Sec. \ref{conclusion}.

\section{Databases and observables} \label{Databases}

The data analyzed in this paper were extracted from InCites, which is a web-based tool including bibliometric indicators about scientific production and citations of institutions and countries. The indicators are generated from the Web of Science (WoS) documents \footnote{The elaborations reported in this paper are based on indicators exported the 2017-09-26 from InCites dataset updated at 2017-09-23 which includes Web of Science content indexed through 2017-07-31}. The indicators at the country level are created based on address criteria using the whole counting method i.e., counts are not weighted by numbers of authors or number of addresses, and all addresses attributed to the papers are counted. As subject area scheme for this study, we use the Essential Science Indicators (ESI) schema which comprises 22 subject areas in science and social sciences and is based on journal assignments. Arts and Humanities journals are not included because their coverage, in terms of publication outputs, is lower compared with other subject areas. Each journal is found in only one of the 22 subject areas and there is no overlap between categories. The Essential Science Indicators are 22 scientific fields categories in which journals are classified. Only one ESI is assigned to each journal, thus the ESI of a paper will be only one, i.e. the ESI of the journal where it is published. Only articles and reviews from Science Citation Index Expanded and Social Science Citation Index are mapped to ESI. Arts \& Humanities, Conference Proceedings Citation Index, and Book Citation Index are excluded. 
Publications in journals such as Nature or Science, which are multidisciplinary, are assigned by Clarivate Analytics to the most pertinent one using the citations of each publication [\onlinecite{Incite}]. 
The ESI fields used to define the disciplinary profiles are: Agricultural Sciences (AGR. SCI.), Biology \& Biochemistry (BIO. \& BIOC.), Chemistry (CHE), Clinical Medicine (CL. MED.), Computer Science (COMP. SCI.), Economics \& Business (ECO. \& BUS.), Engineering (ENG), Environment/Ecology (ENV. ECO.), Geosciences (GEO.), Immunology (IMMU), Materials Science (MAT. SCI.), Mathematics (MATH), Microbiology (MICRO-BIO.), Molecular Biology \& Genetics (MOL. BIO. \& GEN.), Neuroscience \& Behavior (NEUROS. \& BEH), Pharmacology \& Toxicology (PHAR. \& TOX.) Physics (PHYS), Plant \& Animal Science (PLANT. \& AN. SCI.), Psychiatry/Psychology (PSYC.), Social Sciences (SOC. SCI.), Space Science (SPACE SCI.).
Over the period [1980-1988] we analyze the scientific production of 36 countries: 
Argentina (ARG), Australia (AUS), Austria (AUT), Belgium (BEL), Brazil (BRA), Bulgaria (BGR), Canada (CAN), Chile (CHL), China (CHN), Czechoslovakia (CZS), Denmark (DNK), Egypt (EGY), Finland (FIN), France (FRA), Germany (DEU), Great Britain (GBR), Greece (GRC), Hungary (HUN), India (IND), Ireland (IRL), Israel (ISR), Italy (ITA), Japan (JPN), Mexico (MEX), Netherlands (NLD), New Zealand (NZL), Nigeria (NGA), Norway (NOR), Poland (POL), South Africa (ZAF), Soviet Union (USSR), Spain (ESP), Sweden (SWE), Switzerland (CHE), Usa (USA), Yugoslavia (SFRJ).
Over the period [1992-2017] we analyze the scientific production of 50 countries: Argentina (ARG), Australia (AUS), Austria (AUT), Belgium (BEL), Brazil (BRA), Bulgaria (BGR), Canada (CAN), Chile (CHL), China (CHN), Colombia (COL), Czech Republic (CZE), Croatia (HRK), Denmark (DNK), Egypt (EGY), Finland (FIN), France (FRA), Germany (DEU), Great Britain (GBR), Greece (GRC), Hong Kong (HKG), Hungary (HUN), India (IND), Iran (IRN), Ireland (IRL), Israel (ISR), Italy (ITA), Japan (JPN), Malaysia (MYS), Mexico (MEX), Netherlands (NLD), New Zealand (NZL), Norway (NOR), Pakistan (PAK), Poland (POL), Portugal (PRT), Romania (ROU), Russia (RUS), Saudi Arabia (SAU), Singapore (SGP), South Africa (ZAF), South Korea (KOR), Slovenija (SLO), Spain (ESP), Sweden (SWE), Switzerland (CHE) Taiwan (TWN), Thailand (THA), Turkey (TUR), Ukraine (UKR), Usa (USA). 
InCites indicators are quite used. Bornmann and Leydesdorf [\onlinecite{Bornmann}] for instance, using InCites indicators, compare normalized citation impact values calculated for China, Japan, France, Germany, United States, and the UK throughout the time period from 1981 to 2010.
Since the pioneering works, Refs. [\onlinecite{May,King}], the characteristics of the disciplinary structure at the country level has been investigated in many studies [\onlinecite{Yang, Bongioanni, Harzing, Radosevic, Albarran, Lorca,Pinto, issi, Daraio1,Daraio2}]. The analysis of the disciplinary profiles of Eastern Europe contries and Soviet Union and its evolution after the breakup of the Soviet Union has been the subject of several studies [\onlinecite{Markusova, Adams, Guskov, Tregubova, Shashnov}].
Harzing and Giroud [\onlinecite{Harzing}] comparing the profiles of 34 countries across 21 disciplines showed that nations with the fastest increase in their scientific productivity during the periods 1994–2004 and 2002–2012, which tidied up their disciplinary profile towards a more uniform one, then continued relatively unchanging in their well-proportioned disciplinary structures. Almeida et al. [\onlinecite{Almeida}] as Bongioanni et al. [\onlinecite{Bongioanni}] examined disciplinary profiles of European countries across 27 disciplines. Thelwall and Levitt [\onlinecite{Thelwall}] analyzed 26 scientific fields in 25 countries and Pinto and Teixeira [\onlinecite{Pinto}] examined disciplinary profiles of 65 countries over a broad period of time (1980–2016). Different works analyzed the 16  G7 and BRICS countries [\onlinecite{Yang,Shashnov,Yue,Li}] exploring the disciplinary profiles of 45 countries. Recently, the disciplinary profiles of countries from all over the world over the years 2009-2019 for the 22 ESI categories from Clarivate Analytics were investigated in Ref. [\onlinecite{Allik}].

Our observable is the so-called country disciplinary profile, which is the $n_d$-dimensional vector,
\begin{equation}
\textbf{s}_i=(s_i(1),...,s_i(k),...,s_i(n_d)) \label{s},
\end{equation}
where $i$ is the index of country, $s_i(k)=n_i(k)/[\sum_k n_i(k)^2]^{1/2}$ being $n_i(k)$ the number of articles created by the country $i$-th in the $k$-th discipline. The vector $\textbf{s}_i$ has magnitude equal to one, $|\textbf{s}_i|=1$. In the following a $\mu$ suffix will denote a given realization of the variable $\textbf{s}_i$. The index $\mu$ in the present case coincides with a temporal index $t$. Thus ${\textbf{s}_i^{\mu}}$ is a time series related to the $i$-th country. The sampling period of the time series we acquired is one year. The single country can be identified with the node identified by the index $i$ of a network or a graph (world). The number of countries is N, whereas the length of a time series is M. We define furthermore the set of matrix $\textbf{S}=(S(1),...,S(k),...,S(n_d))$. The matrix $S(k)$ has N rows, the $i$-th row is the time series of the $i$-th country. 

\section{Analysis of the cross-correlation matrices} \label{CC}
The analysis of the spectral measure and of eigenvectors statistics of the empirical cross-correlation functions, together with the contrast with RMT, permits to deal with the following points,
\begin{enumerate}[(i)]
\item 
to assess if any and identify what features of the measured cross-correlation functions contain genuine information, discerning them from noise content;
\item 
to analyse the stationarity in time of the genuine information;
\item 
to apply under reliable and verified conditions, as stated in the points (i) and (ii) above, maximum-entropy based inference methods under stationarity condition in order to infer the underlying network of interactions generating the observed variables configurations; 
\item 
to reduce the number of free parameters in the inference procedure, i.e. to lower the rank of the matrix of interactions, thus possibly allowing inference on shorter time series.
\end{enumerate}
In the following we will deal only with the first three points above, while leaving the forth point for further developments and only commenting on it in Sec. \ref{conclusion}.

As a rationale for putting efforts on the first two points we notice that difficulties on applying inference methods based on maximum entropy models is mostly related to the fact that environment conditions can change in time and the resulting cross-correlations may not be stationary. The finite length of the time series, secondly, introduces measurements noise. If a long time series is used in order to circumvent the problem of finite length, the trouble of non-stationarity of the empirical cross-correlations could in place emerge. The contrast between the spectral measure of the empirical correlations with the one (universal) of RMCs, so-called Wishart matrices, permits to identify the non-random components of the measured cross-correlations, which can be thus related to genuine information. As general trend, empirical cross-correlation matrices bringing genuine information are systematically characterized by high-value eigenvalues deviating from the spectral measure of RMCs, which instead matches the  so-called bulk-eigenvalues region of the spectral measure [\onlinecite{RandomMatrix1,RandomMatrix2,Barucca}]. Furthermore, we will analyse the stability in time of the genuine information contained in the time-dependent cross-correlation matrices, thus allowing the use of results holding for stationary states.

\subsection{Cross-correlations and genuine information} \label{crosscorrelations}
In the following we analyse the properties of the pairwise cross-correlation matrix $\textbf{C}=\sum_{k=1}^{n_d}\textbf{C}(k)=\sum_{k=1}^{n_d}\textbf{S}(k)\textbf{S}(k)^T$, where the suffix ``T'' stands for transpose, in the period $[1992-2017]$. The elements of the matrix $\textbf{C}$ are
\begin{equation}
C_{ij}=\frac{1}{M}\sum_{\mu=1}^{M} \textbf{s}_i^{\mu}\cdot \textbf{s}_j^{\mu}=\sum_{k=1}^{n_d}C_{ij}(k)=\sum_{k=1}^{n_d}\frac{1}{M}\sum_{\mu=1}^{M} s_i^{\mu}(k) s_j^{\mu}(k) \label{C}.
\end{equation}
The symbol $\cdot$ stands for a scalar product. 
In order to construct RCM isomorphic to the data cross-correlation matrices we consider a set of $n_d$ independent matrices $\bm{\Sigma}=(\bm{\Sigma}(1),...,\bm{\Sigma}(k),...,\bm{\Sigma}(n_d))$. The matrix $\bm{\Sigma}(k)$ is a $N\times M$ matrix of random elements. The mean value and variance of the entries of each matrix $\Sigma(k)$ is equal to that of $S(k)$. The $\sigma_i^{\mu}(k)$ entries of the matrix $\Sigma(k)$, are then properly normalized so that the vector $\bm{\sigma}_i^{\mu}=(\sigma_i^{\mu}(1),...,\sigma_i^{\mu}(k),...,\sigma_i^{\mu}(n_d)): |\bm{\sigma}_i^{\mu}|=1$.
The RCM $\textbf{R}=\sum_{k=1}^{n_d}\textbf{R}(k)$ has elements
\begin{multline}
R_{ij}= \\ \frac{1}{M}\sum_{\mu=1}^{M} \bm{\sigma}_i^{\mu}\cdot \bm{\sigma}_j^{\mu}=\sum_{k=1}^{n_d}R_{ij}(k)=\sum_{k=1}^{n_d}\frac{1}{M}\sum_{\mu=1}^{M} \sigma_i^{\mu}(k) \sigma_j^{\mu}(k), \label{R}
\end{multline}
\begin{figure*}
\includegraphics[width=.8\textwidth]{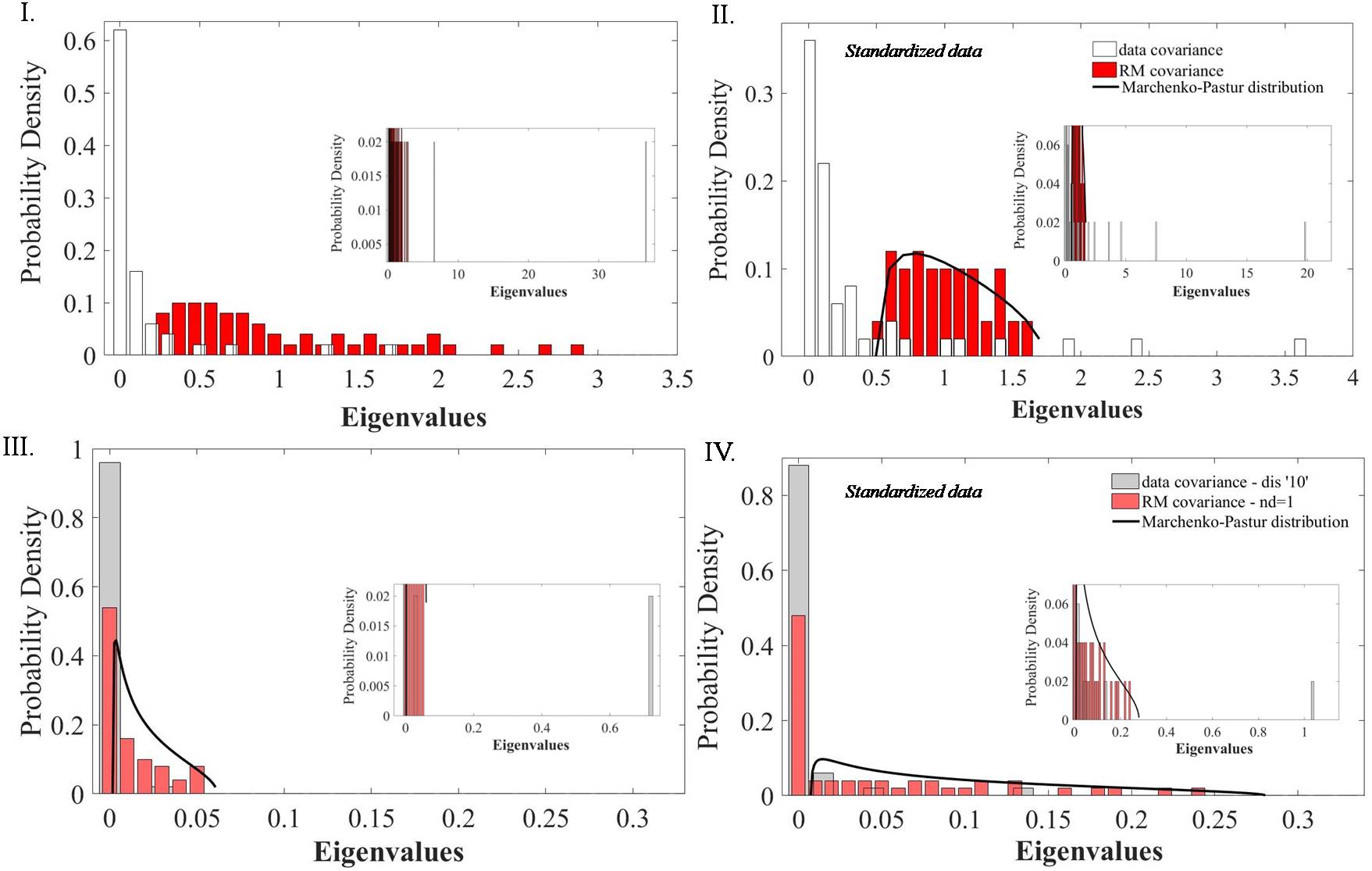}
\caption{\textit{Panel I}. Histogram representation of the eigenvalues distribution of the empirical correlation matrix $\textbf{C}$ of disciplinary profiles in the time interval [1992-2017] (white bars) and of the isomorphic finite-dimensional RCM $\textbf{R}$ (red bars). \textit{Panel II} Histogram representation of eigenvalues distribution of the matrix $\tilde{\textbf{C}}$ (white bars), obtained from standardized data, and $\tilde{\textbf{R}}$ (red bars). The solid curve shows theoretical predictions of RMT. \textit{Panel III} Eigenvalues distribution of $C(k)$ corresponding to MAT. SCI. (light-grey bars) and of $\textbf{R}(k)$ (light-red bars). The solid curve shows the Marchenko-Pastur distribution. \textit{Panel III} Eigenvalues distribution of $\tilde{\textbf{C}}(k)$ corresponding to MAT. SCI. (light-grey bars) and of $\tilde{\textbf{R}}$ (light-red bars). The solid curve shows the Marchenko-Pastur distribution. The insets show a zoom of the large-eigenvalues region of the eigenvalue distribution of empirical covariance matrices.}\label{EigenvaluesFig}
\end{figure*}
Fig. \ref{EigenvaluesFig}, \textit{Panels I}, shows the eigenvalues distribution, $P(\lambda)$, of the data covariance matrix, $\textbf{C}$, defined in Eq. \ref{C} contrasted with the eigenvalues distribution of the RCM, $P_{RCM}(\lambda)$. We also compare the eigenvalues distribution of a single component of $\textbf{C}$, $\textbf{C}(k)$, the index $k$ identifying here MAT. SCI. discipline, with the one of a single term of $\textbf{R}$, $\textbf{R}(k)$, see \textit{Panels III} of Fig. \ref{EigenvaluesFig}. The matrix $\textbf{R}(k)$ is a Wishart matrix, $\textbf{W}$, defined in multivariate statistics [\onlinecite{Mui, Speicher_Text}]. The random matrix $\textbf{R}$ is the sum of Wishart matrices with independent entries, $\textbf{R}=\sum_{k=1}^{n_d}\textbf{W}(k)$. Each matrix $\textbf{W}(k)$ has covariance $\alpha(k)$. The probability distribution of eigenvalues of a Wishart matrix is asymptotically described, in the limit $N \rightarrow \infty$ with $\beta=M/N$ kept constant, by a free Poisson distribution of rate $1/\beta$ and variance $\alpha(k)$. It is known as Marchenko-Pastur distribution, 
\begin{eqnarray}
\rho_{\alpha,\beta}(\lambda)=
\begin{cases}
\tilde{\rho}_{\alpha, \beta}(\lambda), \ \ 0 \leq \beta \leq 1;\\ (1-\frac{1}{\beta})\delta_0+\frac{1}{\beta} \tilde{\rho}_{\alpha, \beta}(\lambda), \ \ \beta > 1 \label{MP}
\end{cases}
\end{eqnarray}
where $\tilde{\rho}_{\alpha, \beta}(\lambda)=\frac{1}{2 \pi \alpha \lambda} \sqrt{4 \beta \alpha^2-[\lambda-\alpha(1+\beta)]^2}$. The symbol $\delta_0$ is the Kronecker delta. We omitted the index $k$ of $\alpha(k)$ in Eq. \ref{MP} to make the notation unclaterd. The measure $\tilde{\rho}_{\alpha, \beta}(\lambda)$ is supported on the interval $[\alpha (1-\sqrt{\beta})^2, \alpha (1+\sqrt{\beta})^2]$. In \textit{Panel II} of Fig. \ref{EigenvaluesFig} it is shown the Marchenko-Pastur distribution (full black line) describing the eigenvalues distribution of $\textbf{W}(k)$. The random matrix $\textbf{R}$ is the sum of selfadjoint matrices with spectral measure given by the Marchenko-Pastur distribution. They are characterized by the same parameter $\beta$ and different variance $\alpha(k)$. The spectral measure of the sum of selfadjoit matrices is given by the free convolution of the spectral measures of each matrix when their size goes to infinity [\onlinecite{Speicher, Speicher_Text}]. Free convolution can be performed by exploiting the so-called R-transform introduced by Voiculescu [\onlinecite{Voiculescu, Speicher_Text}]. In order to gain more insight from RMT predictions we further consider the eigenvalues distribution of cross-correlation matrix, $\tilde{\textbf{C}}$, generated by the set of matrices $\tilde{\textbf{S}}=(\tilde{\textbf{S}}(1),...,\tilde{\textbf{S}}(k),...,\tilde{\textbf{S}}(n_d))$, as $\textbf{C}$ is generated by $\textbf{S}$ following Eq. \ref{C}. The entries of the matrices $\tilde{\textbf{S}}(k)$ satisfy the condition $|\tilde{\textbf{s}}_i^{\mu}|=1$. A further normalization protocol has been applied to $\tilde{\textbf{S}}(k)$: the average value and variance of its entries is zero and one respectively (standardized data). The eigenvalues distribution of $\tilde{\textbf{C}}$ is compared to that of a random matrix $\tilde{\textbf{R}}$ generated by the matrices set $\tilde{\bm{\Sigma}}=(\tilde{\bm{\Sigma}}(1),...,\tilde{\bm{\Sigma}}(k),...\tilde{\bm{\Sigma}}(n_d))$. The entries of each random matrix $\tilde{\bm{\Sigma}}(k)$ have zero average and unitary variance, i.e. $\alpha$ is independent from $k$. The advantage of using the standardized data is that the free convolution of Marchenko-Pastur distributions with same variance $\alpha$ can be easily obtained by using the R-transform, differently from when $\alpha=\alpha(k)$. The probability distribution obtained by the free convolution has the same functional form of Eq. \ref{MP}, but the parameter $\beta$ is replaced by $n_d \beta$. The asymptotic analytical expression of the spectral measure of $\tilde{\textbf{R}}$ is thus computed and shown by a full black line in \textit{Panel II} of Fig. \ref{EigenvaluesFig}. 
In all the analysed cases $P(\lambda)$ shows significant deviations from the corresponding $P_{RCM}(\lambda)$. In particular we notice a deviating behavior of $P(\lambda)$ of $\textbf{C}$ and $\tilde{\textbf{C}}$ in the region of low eigenvalues values, less pronounced in the one-dimensional case (\textit{Panels III} and \textit{IV} of Fig. \ref{EigenvaluesFig}). Furthermore, it is observed the presence of high-value eigenvalues in the empirical cross-correlation matrices (see the insets in Fig. \ref{EigenvaluesFig}) not reproduced neither by the RMT predictions nor by the spectral measure of the finite-dimensional RCMs. 

\begin{figure*}
\includegraphics[width=.8\textwidth]{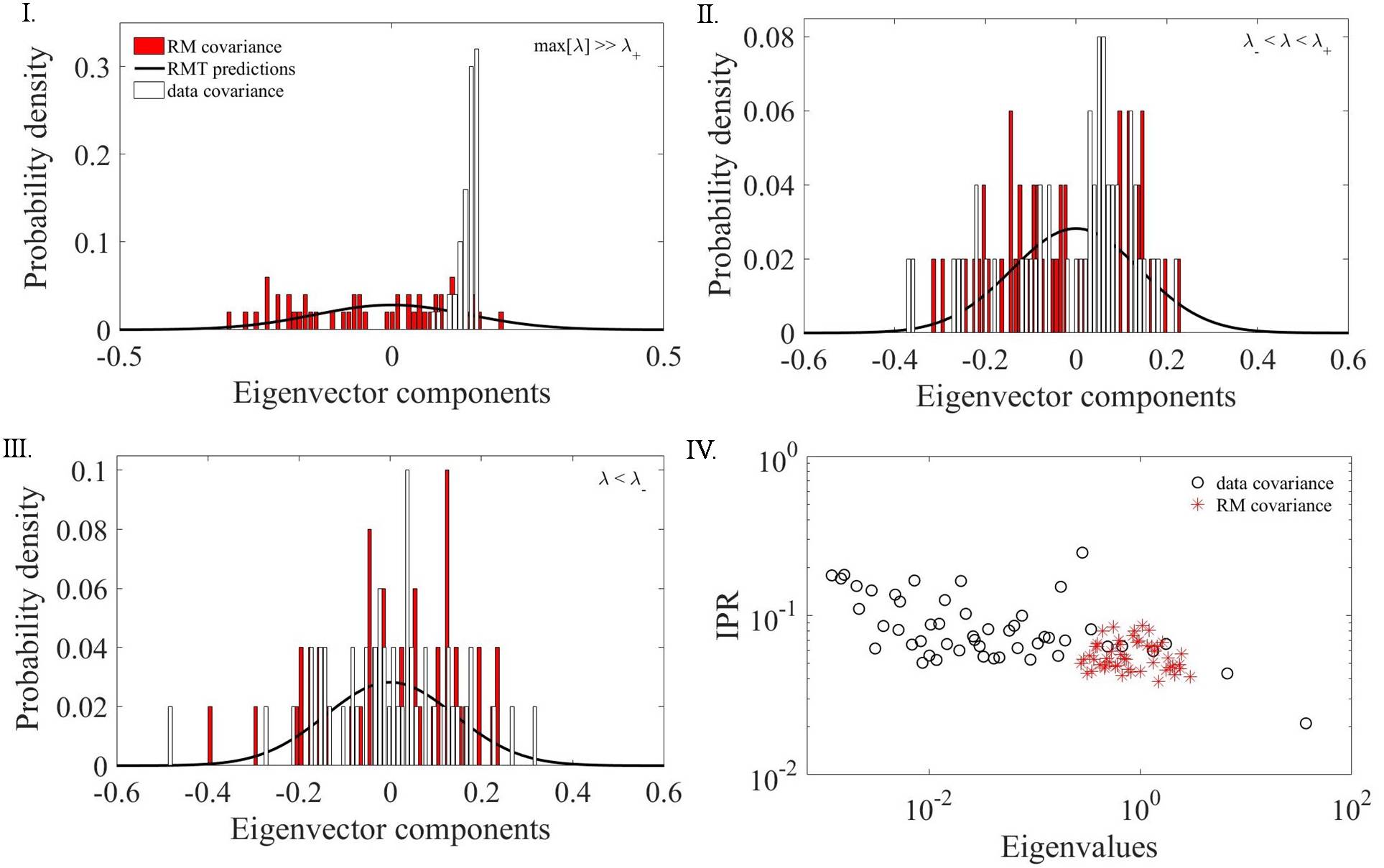}
\caption{\textit{Panel I.} Distribution of eigenvector components (white bars) corresponding to the largest eigenvalue of \textbf{C} ($\lambda \gg \lambda_+$) contrasted with the eigenvector components distribution of test RCM (red bars) and RMT predictions (solid curve). \textit{Panel II.} Distribution of eigenvector components corresponding to a bulk eigenvalue $\lambda$: $\lambda_-<\lambda<\lambda_+$. \textit{Panel III.} Distribution of eigenvector components corresponding to an eigenvalue $\lambda<\lambda_-$. \textit{Panel IV.} Inverse Partecipation Ratio (IPR) as a function of $\lambda$ of the empirical covariance matrix (open circles) and of RCM (stars). }\label{EigenvectorsFig}
\end{figure*}

In order to confirm that the deviating eigenvalues bring genuine information we further analyse the statistics of the corresponding eigenvector components, contrasting it with the one of RCMs and RMT predictions. Fig. \ref{EigenvectorsFig} shows the distribution of components of selected eigenvectors of $\textbf{C}$ corresponding to i) its largest eigenvalue $max[\lambda]$, much larger than the largest of the RCM's eigenvalues, $\lambda_+$, ii) a bulk eigenvalue falling inside the RCM band $[\lambda_-,\lambda_+]$, being $\lambda_-$ the lowest RCM's eigenvalue and iii) an eigenvalue lower than $\lambda_-$. Only in the case of the highest eigenvalue significant deviations from RCM's statistics, which is well described by RMT, are observed. Since no information is contained in an eigenvector of a RCM its N-components distribution is a maximum entropy distribution [\onlinecite{Guhr}], i.e. a Gaussian distribution with zero mean and variance $1/\sqrt{N}$.
In Fig. \ref{EigenvectorsFig}, \textit{Panel IV}, it is finally reported the so-called Inverse Participation Ratio (IPR) of eigenvectors of $\textbf{C}$ and $\textbf{R}$ as a function of the corresponding eigenvalues. The IPR, $I_k$, of the eigenvector $\bm{\xi}_k$ is defined as $I_k=\sum_{i=1}^{N}{\xi_k^i}^4$. It quantifies the reciprocal number of eigenvector components significantly contributing to it. As it is possible to infer from Fig.\ref{EigenvectorsFig}, the IPR of RCM are localized around an average value, $<I>_R=\frac{3}{N}$. 
The IPR of the eigenvectors of $C$ corresponding to the largest eigenvalue significantly deviate from $<I>_R$ and points out a high degree of delocalization of the related eigenvector. The value of the IPR is indeed well represented by $1/N$ showing that all the components contribute equally. This behaviour reveals the delocalized character of the eigenvector, which thus brings information on collective modes of the system [\onlinecite{RandomMatrix1, RandomMatrix2, Barucca,Guhr}]. The IPR associated to the lowest eigenvalues also shows deviations from RMT. Since their IPR is larger than $<I>_R$ the related eigenvectors are, however, localized on only few nodes. It is worth to observe that the two largest components of eigenvectors corresponding to lowest eigenvalues, $\xi_{\lambda(low)}^i$ and $\xi_{\lambda(low)}^j$, always correspond to large cross-correlation terms, $C_{ij}$.

\begin{figure*}
\includegraphics[width=1\textwidth]{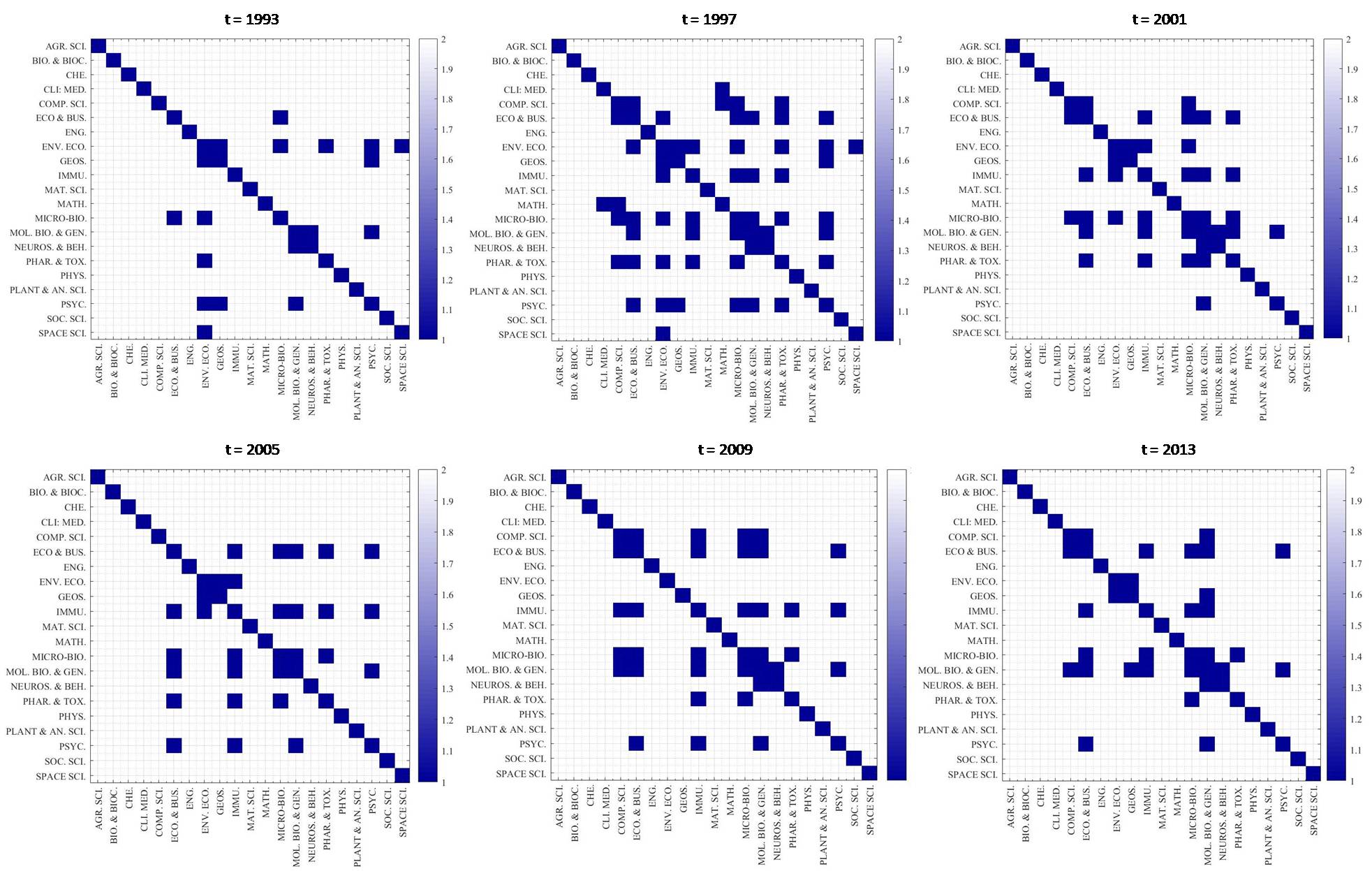}
\caption{Number of eigenvalues different from zero in a matrix obtained by the sum of two empirical cross-correlation matrices related to two different disciplines, $\textbf{C}(k)+\textbf{C}(l)$. Each cross-correlation matrix is calculated for a single time t.}\label{Dip_Disc}
\end{figure*}

Finally, we observe that both the single-discipline $\textbf{C}(k)$ and $\textbf{R}(k)$ have a finite number of null eigenvalues, see Fig. \ref{EigenvaluesFig}, \textit{Panels III/IV}. According to their own definition $\textbf{C}(k)$ and $\textbf{R}(k)$ have rank $N-M$. The existence of null eigenvalues can hamper the application of mean-field approximation to infere the couplings, since it requires inverting the covariance matrix [\onlinecite{MF1,Decelle, DirectCouplings}]. In parallel, in optimization problems, such as the maximization of the Likelihood function described in the following or minimization of the Chi-square function, finite size-effects of the time series can make the number of observed configurations much smaller than the number of free parameters and lead, e.g., to negative-valued averaged Chi-square. In the vector case, if the components of the vector $\textbf{s}_i$ are uncorrelated among themselves the number of degree of freedom of the time series of $\textbf{s}_i$ is increased restoring, e.g., a positive value of the averaged Chi-square.  In the Supplemental Material they are shown the eigenvalues of $\textbf{C}$ and $\textbf{C}(k)$. The analysis of low-value eigenvalues of the empirical cross-correlation matrix can also represent a valuable tool to fix the optimal classification scheme of disciplines in the context of our inference problem. If a very dense classification scheme is used, with a very large number of disciplines, correlations among disciplines production will arose introducing redundant information without increasing the degrees of freedom. If a poorly resolved classification scheme is adopted, matters related to small number of degrees of freedom will show up. Correlations among a couple of disciplines is highlighted by considering the number of eigenvalues different from zero of the matrix sum of two single-discipline cross-correlation matrices calculated for a given configuration. Dependency between the production of the two disciplines is present if the number of eigenvalues different from zero is one, otherwise independency can be assumed. Fig. \ref{Dip_Disc} shows for all the possible couple of disciplines considered in this work the number of eigenvalues different from zero, thus pointing out pair-wise correlations among disciplines.
%
\begin{figure*}
\includegraphics[width=.65\textwidth]{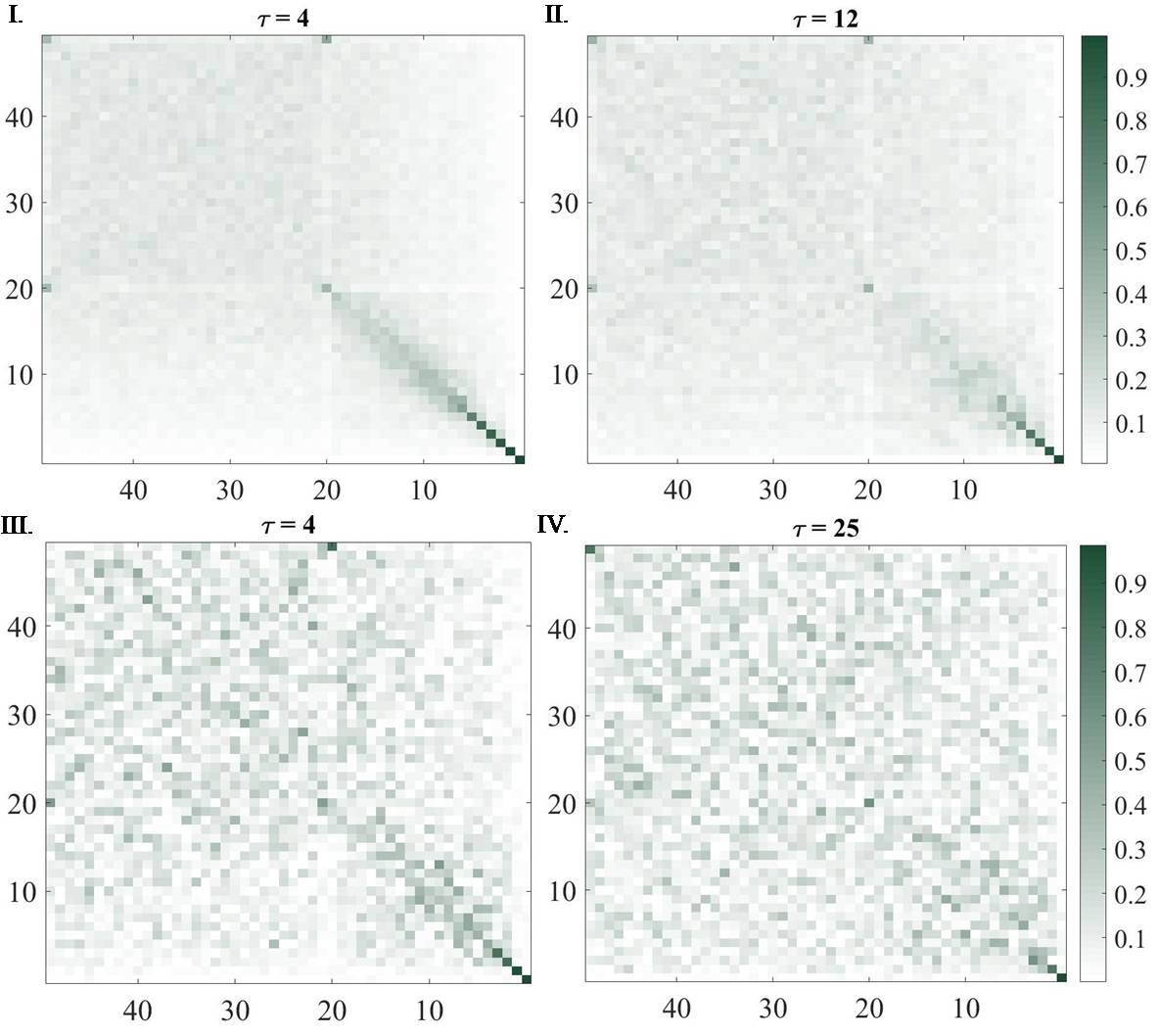}
\caption{\textit{Panels I./II.} $\overline{O}(\tau)$ for two different delay time $\tau$. \textit{Panels III./IV.} $O(t=1,\tau)$ for two different delay time $\tau$.}\label{O_t}
\end{figure*}
\begin{figure*}
\includegraphics[width=1.0\textwidth]{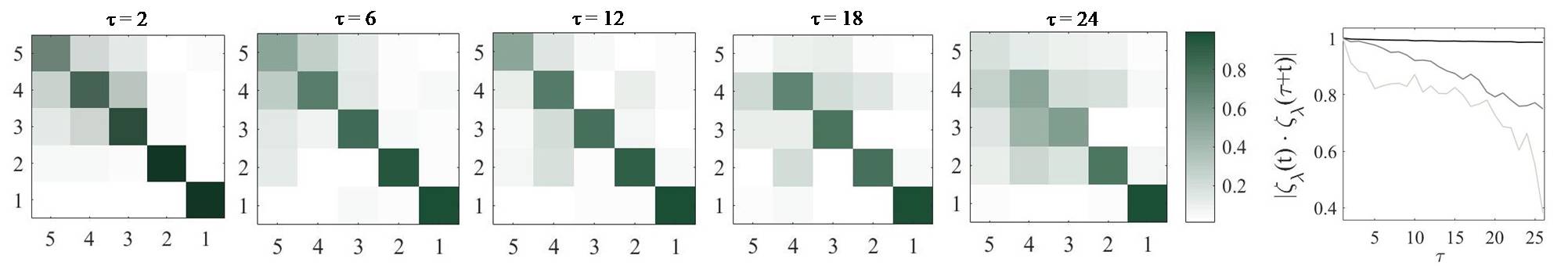}
\caption{$O(t=1,\tau)$ for only the eigenvectors corresponding to the five largest eigenvalues at different delay time $\tau$. The autocorrelation, $|\zeta_{\lambda}(t) \cdot \zeta_{\lambda}{t+\tau}|$ of the eigenvectors corresponding to the first three largest eigenvalue is shown for $t=1$ as a function of $\tau$.}\label{O_t2}
\end{figure*}
\subsection{Stationarity properties}
We verified that the empirical cross-correlations bring genuine information, pointing out how this information is enclosed in the largest eigenvalues of $C$ and corresponding eigenvectors. We analyze the stability in time of such eigenvectors. Since deviations from RCM outcomes imply genuine correlations \textit{de facto} related to the underlying interactions network, they should show some degree of stability in the time interval used to compute $\textbf{C}$ if the interactions network remains stable and if the system is in a stationary state. We define: i) the overlap matrix $\textbf{O}(t,\tau)=\textbf{V}(t)\textbf{V}^T(t+\tau)$, where $\textbf{V}(t)$ is a matrix whose columns are the eigenvectors of the correltaton matrix at time $t$ (notice that for sake of clarity we substitue here the index $\mu$ with the index t) corresponding to eigenvalues sorted in ascending order; ii) the average overlap matrix $\overline{\textbf{O}}(\tau)=<\textbf{O}(t,\tau)>_t$, i.e. the overlap matrix $\textbf{O}(t,\tau)$ averaged over all the starting time $t$ included in the measured time interval. The entries of the matrix $\overline{\textbf{O}}(\tau)$ for selected values of $\tau$ and of $\textbf{O}(t,\tau)$ for selected values of $\tau$ and $t=1$ are displayed respectively in \textit{Panels I/II} and \textit{III/IV} of Fig. \ref{O_t}. If all the eigenvectors of the matrix $C(t)$ were non-random and stationary, both $\overline{\textbf{O}}(\tau)$ and $\textbf{O}(t,\tau)$ would be diagonal with entries equal to one. As shown in Fig.\ref{O_t} this condition is approximately satisfied only for eigenvectors corresponding to $\lambda \gg \lambda_{+}$. In particular the eigenvector related to the largest eigenvalue remains stable for all the period under consideration, see \textit{Panel IV} of Fig. \ref{O_t}, . It is interesting to note that a certain degree of stability is also shown by the eigenvector corresponding to the lowest eigenvalue.

\section{\textbf{Inference Method}} \label{method}
\subsection{\textbf{Maximum-entropy estimates of second-order marginal and definition of the Likelihood function}} \label{entropy}

The Shannon theorem [\onlinecite{Shannon}] states that the entropy $S$ defined in statistical mechanics is a measure of the 'amount of uncertainty' related to a given discrete probability distribution $p$ of variables configuration $\textbf{s}=\{\textbf{s}_1, ..., \textbf{s}_N\}$,
\begin{eqnarray}
S[p]=-K \sum_{\{ \textbf{s} \}} p(\{ \textbf{s}\}) \log [p(\{ \textbf{s}\})]. \label{entropy}
\end{eqnarray} 
$K$ is a positive constant, hereafter taken equal to one. This quantity is positive and additive for independent sources of uncertainty.
In making inference on the basis of partial available information, the probability which maximizes the 'amount of uncertainty' or entropy subject to whatever is known [\onlinecite{Jaynes}] has to be used. 
Since the empirical expectation values are known, formally this means that $p(\{\textbf{s}\})$ is found as a solution of a constrained optimization problem, i.e. the entropy of the distribution should be maximized subject to conditions that enforce the expectation values to coincide with the empirical ones. One refer to the quantities
whose averages are constrained as 'features' of the
system. 
As emphasized in Sec. \ref{introduction}, we only take into account pairwise interactions, neglecting higher order of interactions. Such a choice could be \textit{a posteriori} validated in the case where an empirical probability distribution of configurations $\{\textbf{s}\}$ can be obtained from the data, i.e. when the number of acquisitions is large enough [\onlinecite{PairEntropy1,PairEntropy2}]. This is not the case for the data we are handling. We, however, shortly describe in the following the protocol in order to assess this. The content of information enclosed in a given order interaction (pairwise, triplet, and so on) can be quantified by defining the maximum entropy related to the marginal of order $k$, where $k=1,2,..., N$ being $N$ the total number of observables. Given a joint probability distribution $p(\{\textbf{s}_1,...,\textbf{s}_N \})$ the marginal of order k is then defined as $p_{k}(\{\textbf{s}_1,...,\textbf{s}_k \})=\sum_{\textbf{s}_j \neq \textbf{s}_1,...,\textbf{s}_k}p(\{\textbf{s}_1,...,\textbf{s}_N \})$. The marginal $p_k$ can be also defined as the maximum entropy distributions that are consistent with the $k$-th order correlations. The marginal of order $N$ corresponds to the exact distribution of the $N$ correlated variables, whereas the marginal of order $1$ states for the distribution of $N$ independent variables. The entropies related to marginals of a given order, $S_k$, decreases monotonically by increasing $k$ towards the true entropy, $S_N$. The connected information or entropy difference, $I_k=S_{k-1}-S_k$, represents the amount by which the maximum possible entropy of the system decreases when one goes from including marginals of order $k-1$ to including also marginals of order $k$, thus providing a characterization of the relative importance of various orders of interaction [\onlinecite{PairEntropy1,PairEntropy2, Roudi}]. The multi-information $I_N=S_1-S_N$ instead quantify the total amount of correlation in the network, independent of whether it arises from pairwise or higher order interactions. The contrast between $I_N$ and $I_2$ permits thus to assess if the pairwise interaction model provide an effective description of the system.

Observing variables only in pairs, the optimization problem reduces to 
\begin{eqnarray}
Max_{p(\{ \textbf{s}_i\})}S[p(\{\textbf{s}_i\})], \label{maxentro}
\end{eqnarray} 
with the constraints
\begin{eqnarray}
&& \sum_{\{ \textbf{s}_i\}}p(\{ \textbf{s}\})=1 \label{constr1} , \\ && <\textbf{s}_i \cdot \textbf{s}_j>_{p(\{ \textbf{s}\})}=\frac{1}{M}\sum_{\mu=1}^{M} \textbf{s}_i^{\mu}\cdot \textbf{s}_j^{\mu}. \label{constr2}
\end{eqnarray}
The features are $f_{ij}=\textbf{s}_i \cdot \textbf{s}_j$. The sum is over all possible configurations in the phase space. Eq. \ref{maxentro} with the constraints \ref{constr1} and \ref{constr2} is solved by using the Lagrange multipliers $\lambda_0$, $\{\lambda_{ij}\}$,
\begin{eqnarray}
p(\{ \textbf{s}\}|\{ \lambda \})=\frac{1}{Z}e^{-\frac{1}{2}\sum_{i \neq j}\lambda_{ij}\textbf{s}_i \cdot \textbf{s}_j}, \label{p}
\end{eqnarray}
with $Z=\sum_{\{ \textbf{s}\}}e^{-\sum_{i \neq j}\lambda_{ij}\textbf{s}_i \cdot \textbf{s}_j}$. The constants $\{ \lambda_{ij}\}$ are obtained by the constraint \ref{constr2}. The probability distribution, Eq. \ref{p}, is a Boltzmann distribution related to a generalized Heisenberg model with pairwise interactions $\lambda_{ij}$, $n_d$-dimensional spin variables and zero external field \footnote{This does not represent a loose of generality. Couplings and magnetic field can indeed be changed together without changing the sum in the exponent of Eq. \ref{p} (gauge invariance).}. 

It is possible to reformulate the problem of maximizing the entropy subject to constraint on the expectation values of pairwise correlation functions as searching the maximum of the so-called Likelihood function within the class of Boltzmann probability distribution related to the generalized Heisenberg model with multidimensional spin variables.
The Likelihood function is introduced in the context of Bayesian inference [\onlinecite{Barber1}]. Given the set of variables $\{\textbf{s}\}$ and the set of data $\{\textbf{s}^{\mu}\}$, it is assumed that i) each realization of the set $\{ \textbf{s} \}$, $\{\textbf{s}^{\mu}\}$, is drawn independently, ii) the data have been generated by a (known) model, which depends on the set of (unknown) pairwise parameters $\{ J \}$, with generic element $J_{ij}$. One aims to find the optimal values of $ \{ J \}$, which maximize the conditional probability [\onlinecite{Barber1}] 
\begin{multline}
p(\{J\}|\{\textbf{s}\})=\\\frac{p(\{\textbf{s}\}|\{J\})p(\{ J\})}{p(\{ \textbf{s} \})}=\frac{p(\{\textbf{s}\}|\{J\})p(\{ J\})}{\int_{\{J\}}p(\{\textbf{s}\}|\{J\})p(\{ J\})}. \ \ \label{lk}
\end{multline}
The probability $p(\{J\}|\{\textbf{s}\})$ is called \textit{posterior}, $p(\{ J\})$ \textit{prior}, $p(\{ \textbf{s} \})$ \textit{evidence} and $p(\{\textbf{s}\}|\{J\})$ \textit{Likelihood}. If the prior is the uniform distribution, as we assume here, the most probable a posteriori set of variable is, as a consequence of Eq. \ref{lk}, the one that maximizes the Likelihood function, 
\begin{multline}
p(\{ \textbf{s}\}|\{ J\})=\\\frac{1}{Z(\{ J\})}e^{-H(\{ \textbf{s}\}|\{ J\})}=\frac{1}{Z(\{ J\})}e^{-\frac{1}{2} \sum_{i \neq j}^{1,N} J_{ij} \textbf{\textbf{s}}_i \cdot \textbf{\textbf{s}}_j}. \label{mn_1}
\end{multline} 
The set $\{ J\}$ with generic elements $J_{ij}$ identify the pairwise interactions of the Heisenberg model. The partition function $Z(\{ J\}) =\sum_{\{\textbf{s}\}} e^{-H(\{\textbf{s}\} | \{ J\})}$. 
The Hamiltonian or cost function is
\begin{equation}
H(\{\textbf{s}\} | \{ J\}) = \frac{1}{2} \sum_{i \neq j}^{1,N} J_{ij} \textbf{\textbf{s}}_i \cdot \textbf{\textbf{s}}_j, \label{cost}
\end{equation}
Given the hypothesis of independence of the data set $\{\textbf{s}^{\mu}\}$ and the functional form of the Likelihood function in Eq. \ref{mn_1}, the Log-Likelihood function, $l(\{J\})$ is given by [\onlinecite{Barber}]
\begin{multline}
l(\{J\}) =\\ \log(L(\{J\} ) = \sum_{\mu=1}^{M} - H(\{\textbf{s}^{\mu} \}| \{J\}) - M\log (Z (\{J\})). \label{log_like_1}
\end{multline}
It is immediate to verify as shown in the following that the maximum of the Log-Likelihood function is given by Eq. \ref{p}, once the constants $\lambda_{ij}$ have been identified with the pairwise interaction parameters $J_{ij}$. The gradient of the Log-Likelihood function is
\begin{eqnarray}
\frac{\partial}{\partial J_{ij}} l(\{ J\})= \frac{1}{2}M \big[C_{ij}- <\textbf{s}_i\cdot\textbf{s}_j>_{\{ J\}}\big], \label{gralike}
\end{eqnarray}
where $< \ >_{\{ J\}}$ states for ensemble average calculated with the probability distribution $p(\{ s \}|\{J \})$, Eq. \ref{mn_1}, and parameters $\{ J\}$.
Under the hypothesis of ergodicity of the system under account, when the ensemble average is calculated with the 'true' set of parameters, i.e. the one whose associated distribution has actually generated the data, in the limit $M \rightarrow \infty$, $C_{ij} \rightarrow <\textbf{s}_i\cdot\textbf{s}_j>_{\{J\}}$ and $\frac{\partial}{\partial J_{ij}} l(\{ J\}) \rightarrow 0$. The maximum of $l(\{ J \})$ in the limit $M \rightarrow \infty$ is thus obtained for those values of $\{ J\}$ which generated the correlations $C_{ij}$. The optimal value of $J_{ij}$ are thus those for which Eq. \ref{gralike} is equal to zero, in agreement with the constraint \ref{constr2} determining $\{\lambda\}$. The hypothesis of ergodicity is assumed without further validation. 

The Ising or Heisenberg model, have been largely exploited in different fields, beyond the original application to magnets in statistical physics, e.g. in image processing, spatial statistics [\onlinecite{image_1,image_2,image_3}] and social networks [\onlinecite{social}]. It is nowever worth to observe that by exploiting the Shanon theorem, the Ising or Heisenberg model does not arise from any specific hypotheses about the network but it comes out as the least-structured model consistent with the measured pairwise correlations. 

\subsection{\textbf{Pseudo-Likelihood approach for the generalized  Heisenberg model with $n_d$-dimensional spin variables}} \label{Heisenberg}
While the definition of the Likelihood function has strong theoretical roots, the realization of an optimization algorithm able to draw the optimal $\{ J\}$ is hindered by the general intractability of computing the partition function and its gradient. Maximum Pseudo-Likelihood estimation avoids this computational issue entirely by optimizing a different objective function: the Pseudo-Likelihood, which has the advantage to be maximized in polynomial time [\onlinecite{B}]. The Pseudo-Likelihood function is based on the local conditional Likelihood at each node of the network [\onlinecite{Barber1,issi}].
The local conditional probability (single-variable Pseudo-Likelihood) at the $i$-th node is
\begin{eqnarray}
p(\textbf{s}_i | \{\textbf{s}_{\backslash i} \},\{ J\})=\frac{1}{Z_i(\{ J\})}e^{-H_i (s_i | \{\textbf{s}_{\backslash i} \},\{ J\})}, \label{pseudo_sing}
\end{eqnarray} 
where $\textbf{s}_{\backslash i}$ indicates the set of all input-variables except the $i$-th.The local hamiltonian $H_i (s_i | \{\textbf{s}_{\backslash i} \},\{ J\})=-\textbf{s}_i \cdot [\frac{1}{2} \sum_{i \neq j}^{1,N} J_{i,j} \textbf{s}_j]$ and the local partition function is $Z_i(\{ J\})=\sum_{\{\textbf{s}_i\}}e^{-H_i (s_i | \{\textbf{s}_{\backslash i} \},\{ J\})}$. By defining $L'(\textbf{s}_i|\{\textbf{s}\}_{\ \_i|\{ J\}})=\log[p(\textbf{s}_i | \{\textbf{s}_{\backslash i} \},\{ J\})]$, the Pseudo-Likelihood function is 
\begin{eqnarray}
l'(\{ J\})= \sum_{\mu=1}^{M} \sum_{i=1}^N L'^{\mu}(\textbf{s}_i|\{\textbf{s}\}_{\ \_i|\{ J\}}). \label{logLi}
\end{eqnarray} 
It is possible to show that the Pseudo-Likelihood maximization is exact (i.e. it is maximized by the same set of parameters than the Likelihood function) in the limit of infinite sampling [\onlinecite{hyv, Aurell}], as discssed in the following.
Because in Eq. \ref{cost} anti-symmetric piece with respect to the index $i$ and $j$ in the Hamiltonian would cancel, the interactions $J_{ij}$ can be chosen symmetric and the Hamiltonian rephrased as $H=-\sum_{i<j}^{1,N}J_{ij} \textbf{s}_i \cdot \textbf{s}_j$. The Hessian of both the Likelihood and Pseudo-Likelihood functions is thus a triangular matrix. The diagonal elements of the Hessian of the Likelihood function, e.g., are $\frac{\partial^2}{\partial J_{ij}^2} l(\{ J\})=<\textbf{s}_i\cdot\textbf{s}_j>_{\{ J\}}^2-(<\textbf{s}_i\cdot\textbf{s}_j>_{\{ J\}})^2$. The latter quantities, apart from some pathological cases where they could be zero, are negative. Since the eigenvalues of a triangular matrix are the entries on its main diagonal, the Likelihood function is strictly concave. Similarly, the Pseudo-Likelihood function is also concave. 
The gradient of the Log-Pseudo-Likelihood function with respect to the parameter $J_{ij}$ is
\begin{eqnarray}
\frac{\partial}{\partial J_{ij}} l'(\{ J\})= \frac{1}{2}M \big[C_{ij}- <\textbf{s}_i\cdot\textbf{s}_j>_{i,\{ J\}}\big], \label{grapseudo}
\end{eqnarray}
where $<\ \ >_{i,\{J\}}$ states for ensemble average calculated for the probability distribution $p(\textbf{s}_i | \{\textbf{s}_{\backslash i} \},\{ J\})$.
It is possible to rephrase the gradient of the Log-Likelihood function, Eq. \ref{gralike}, obtaining
\begin{eqnarray}
\frac{\partial}{\partial J_{ij}} l(\{ J\})= \frac{1}{2}M \big[C_{ij}-<<\textbf{s}_i\cdot\textbf{s}_j>_{i,\{ J\}}>_{\{J\}}\big], \label{gra2}
\end{eqnarray}
By comparing Eq. \ref{grapseudo} and \ref{gra2} it is possible to infer that in the limit $M \rightarrow \infty$ (infinite sampling): i) both the gradients go to zero, ii) $\frac{\partial}{\partial J_{ij}} \lambda(\{ J\}) \rightarrow \frac{\partial}{\partial J_{ij}} l(\{ J\}) $. Because of the concavity of both functions this finally proves the exact maximization of the Pseudo-Likelihood function for $M \rightarrow \infty$.

In the case of a $n_d$-dimensional Heisenberg model with interaction parameters not restricted to nearest neighbor nodes, the partition function and the gradient of the Pseudo-Likelihood function can be calculated analytically, thus facilitating the computational solution of the inference problem through steepest descent method. The partition function is given by
\begin{eqnarray}
Z_i=\int_{-\infty}^{\infty} d\textbf{s}_i e^{\textbf{s}_i \cdot \textbf{A}_i} \delta(s_i-1),
\end{eqnarray}
with $\textbf{A}_i=-\frac{1}{2}\sum_{j=1}^N J_{ij}\textbf{s}_j$, $s_i=|\textbf{s}_i|$ and $A_i=|\textbf{A}_i|$. Introducing polar coordinates ($z$ axis parallel to $\textbf{A}_i$), so that $\textbf{s}_i \cdot \textbf{A}_i=s_iA_i cos \theta$, it is
\begin{widetext}
\begin{multline}
Z_i=\int_{0}^{\infty}d{s}_i \ {s}_i^{n_d-1} \delta({s}_i-1) \ \omega_{n_d-2}\int_0^{\pi} e^{{s}_i{A}_icos \theta} (\sin \theta)^{n_d-2} d \theta = \omega_{n_d-2} \int_{0}^{\pi} e^{A_icos \theta} (\sin \theta)^{n_d-2} d \theta = \\ A_i^{-\frac{n_d-2}{2}}(2\pi)^{\frac{n_d}{2}}J_{\frac{n_d-2}{2}}(iA_i),
\end{multline}
\end{widetext}
where $\omega_n$ is the area of the unit sphere in $n$-dimensional space, being $\omega_n=\frac{(2 \pi)^{\frac{n+1}{2}}}{\Gamma(\frac{n+1}{2})}$ with $\Gamma(n)$ the Gamma function. The Bessel function of order $n$, $J_n(t)$ is defined as $J_n(t)=\frac{t^n}{(2 \pi)^{n+1}}\omega_{2n}\int_0^{\pi}e^{-it cos \theta } (sin \theta) ^{2n} d \theta$. 
To obtain an exact expression of the gradient of the Pseudo-Likelihood function from Eq.\ref{grapseudo} it is needed to calculate $ <\textbf{s}_i\cdot\textbf{s}_j>_{i,\{ J\}}$. We find
\begin{multline}
<\textbf{s}_i \cdot \textbf{s}_j>_{i,\{ J\}}=\frac{1}{Z_i} \int_{-\infty}^{\infty} d \textbf{s}_i e^{\textbf{s}_i \cdot \textbf{A}_i} \textbf{s}_i \cdot \textbf{s}_j \delta ({s}_i-1)=\\ \frac{1}{Z_i}\sum_{\alpha=1}^{n_d} \frac{\partial}{\partial A_i^{\alpha}} Z_i s_j^{\alpha}=\frac{J_{\frac{n_d+1}{2}}(iA_i)}{J_{\frac{n_d-2}{2}}(iA_i)}\hat{\textbf{A}}_i \cdot \textbf{s}_j,
\end{multline}
where $\hat{\textbf{A}_i}=\frac{\textbf{A}_i}{A_i}$. In the case that the dimension $n_d$ is odd the Bessel function can be analytically expressed in terms of elementary functions, obtaining
\begin{multline}
Z_i=\omega_{n_d-2} 2 \sum_{k=0}^{\nu}\Big[ \frac{\nu !}{k! (\nu-k)!} (-1)^k \\\sum_{l=0}^k \frac{(2k)!}{(2k-2l)!} \frac{1}{{A_i}^{2l}} \big( \frac{\sinh{A_i}}{A_i}-\frac{\cosh{A_i}}{{A_i}^2}(2k-2l)\big) \Big],
\end{multline}
where $\nu \in N:n_d-2=2\nu+1$. Furthermore
\begin{widetext}
\begin{multline}
\frac{\partial}{\partial A_i^{\alpha}} Z_i=\omega_{n_d-2}2\sum_{k=0}^{\nu} \frac{\nu !}{k! (\nu-k)!} (-1)^k \sum_{l=0}^k \frac{(2k)!}{(2k-l)!} \frac{A_i^{\alpha}}{{A_i}^{2l+1}}\big( \sinh{A_i}-\frac{\cosh{A_i}}{A_i}(2k-l-1)-\\ \frac{\sinh{A_i}}{{A_i}^2}(2k-l-1)+2\frac{\cosh{A_i}}{{A_i}^3}(2k-l) \big). \label{grapseudodisp}
\end{multline}
\end{widetext}

The maximization of the Log-Log-Pseudo-Likelihood functions has been performed by means of the MATLAB fminunc package [\onlinecite{Matlab}] by selecting a trust-region optimization algorithm. A \textit{l2}-regularizer (parameter 0.13) was used  [\onlinecite{Ravikumar, Tyagi}].

\section{\textbf{The inferred interactions network}} \label{network}
\subsection{\textbf{Assesment of the inferred interactions \textbf{J}}} \label{J_RMT}
\begin{figure*}
\includegraphics[width=0.75\textwidth]{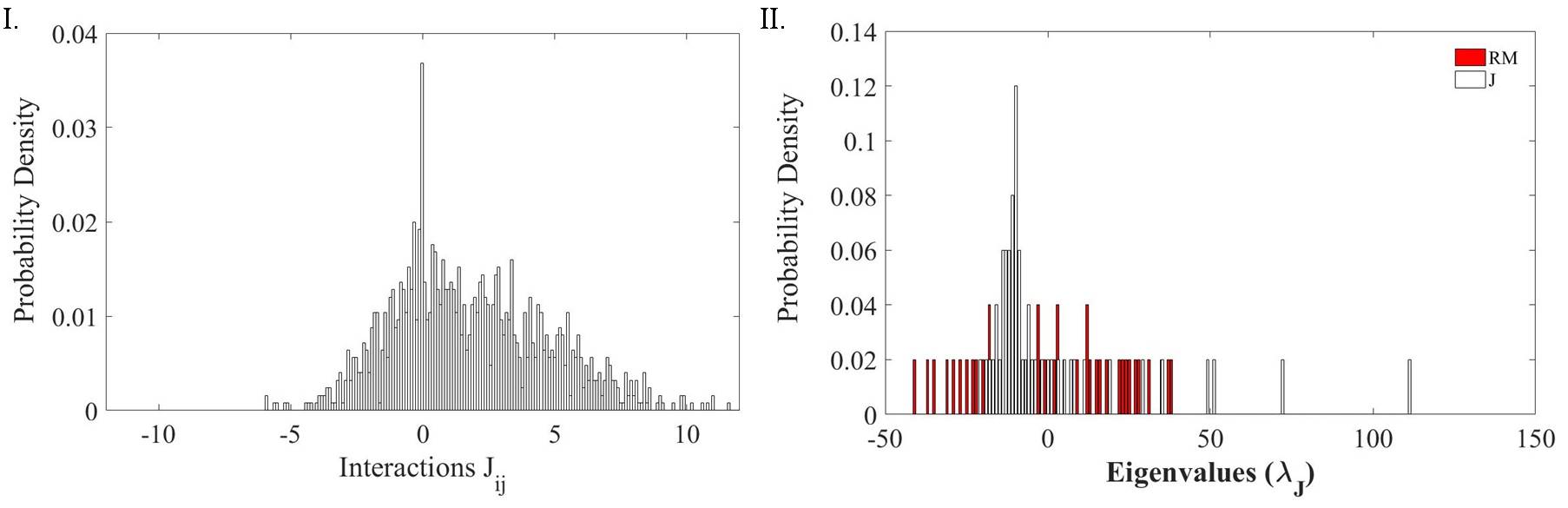}
\caption{\textit{Panel I.} Histogram representation of the distribution of $J_{ij}$. \textit{II.} Histogram representation of the eigenvalues distribution of the interactions matrix $\textbf{J}$ (white bars). The red bars show the eigenvalues distribution of an isomorphic Gaussian random matrix with same average and variance of $\textbf{J}$ and zero diagonal elements.}\label{ProbJ}
\end{figure*}

\begin{figure*}
\includegraphics[width=0.75\textwidth]{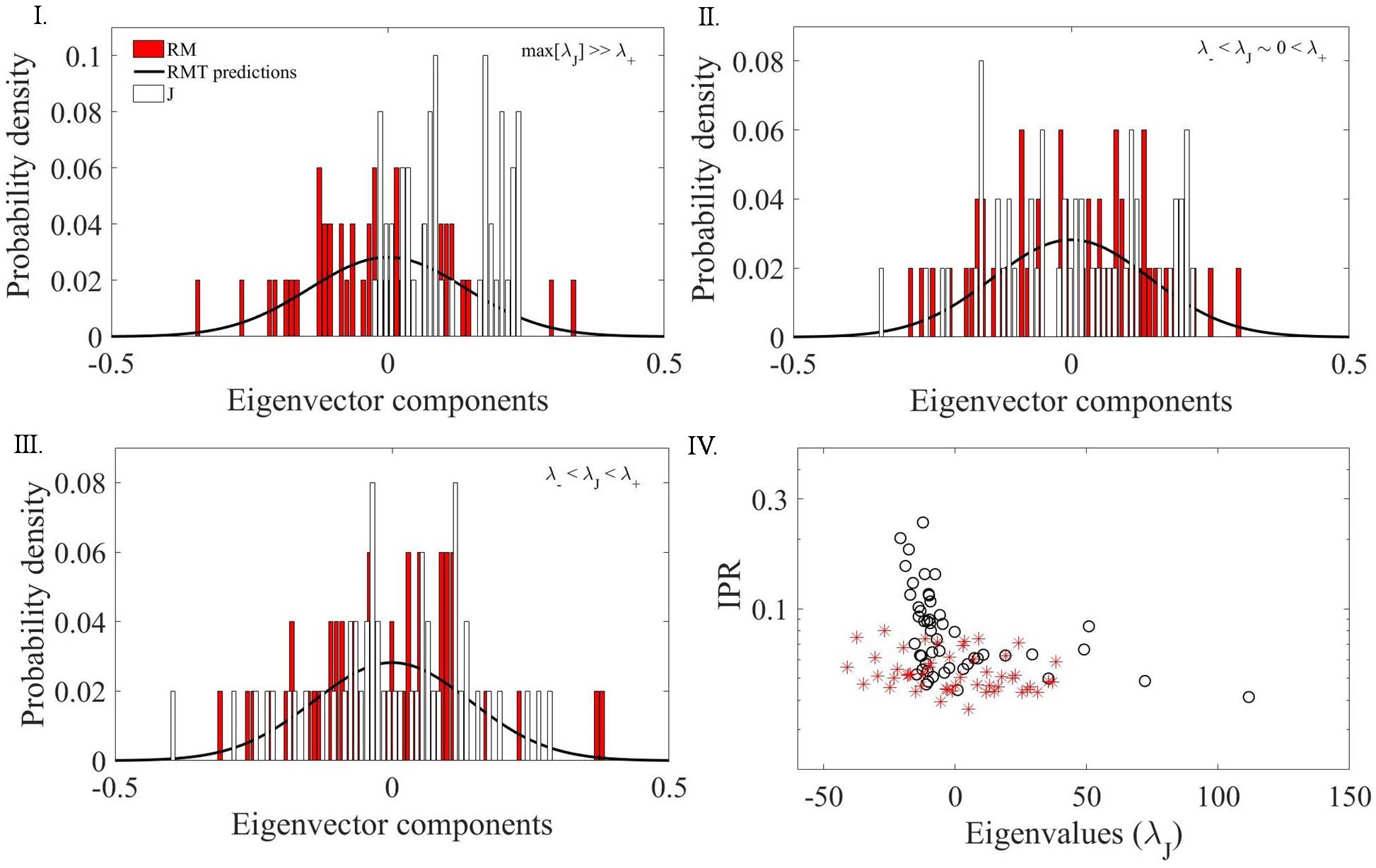}
\caption{\textit{Panel I.} Distribution of eigenvector components (white bars) corresponding to the largest eigenvalue of the interactions matrix $\textbf{J}$, $\lambda_{J} \gg \lambda_+$, contrasted with the eigenvector distribution of a test Gaussian random matrix (red bars) and RMT prediction (solid line). \textit{Panel II.} Distribution of eigenvector components corresponding to a bulk eigenvalue $\lambda_J$: $\lambda_-<\lambda_J<\lambda_+$ and $\lambda_J \sim 0$. \textit{Panel III.} Distribution of eigenvector components corresponding to an eigenvalue $\lambda_J$: $\lambda_-<\lambda_J<\lambda_+$. \textit{Panel IV.} Inverse Participation Ratio (IPR) as a function of $\lambda_{J}$ of $\textbf{J}$ (open circles) and a test Gaussian random matrix (stars). }\label{EigJ}
\end{figure*}

In order to attest the consistency of the inference protocol as well as to analyse the content of information contained in the set of inferred interactions, the spectral measure and the eigenvectors of $\textbf{J}$ are contrasted with corresponding RCMs. Fig.\ref{ProbJ}, \textit{Panel I}, shows a histogram representation of the probability distribution of the values of $J_{ij}$. 
We define a random matrix, \textbf{R}', whose entries are extracted by a normal distribution. Its mean value and variance are set to the ones of $\{ J \}$, the diagonal elements are set to zero and the matrix is furthermore made symmetric. 
The spectral measure of $\textbf{J}$, displayed in \textit{Panel II} of Fig. \ref{ProbJ}, is compared to the one of \textbf{R}'. Similarly to the case of the empirical cross-correlations matrices, the spectral distribution of \textbf{J} covers larger values than the one of \textbf{R}'. 
The comparison between the probability distribution of the eigenvectors ($\bm{\zeta}_k$) components related to selected eigenvalues ($\lambda_J$) of \textbf{J} and \textbf{R}' confirms the non-random character of the eigenvector associated to the largest eigenvalue, see Fig. \ref{EigJ}, \textit{Panels I-III}. The IPR of the eigenvectors of \textbf{J} and \textbf{R}', displayed in \textit{Panel IV} of Fig.\ref{EigJ} as a function of $\lambda_J$, further reveals a deviation from RMT results for largest and smallest eigenvalues. As in the case of the cross-correlations, the eigenvectors corresponding to largest eigenvalues have a delocalized character, whereas the ones associated to lowest eigenvalues show strong localization. 
\begin{figure*}
\includegraphics[width=0.75\textwidth]{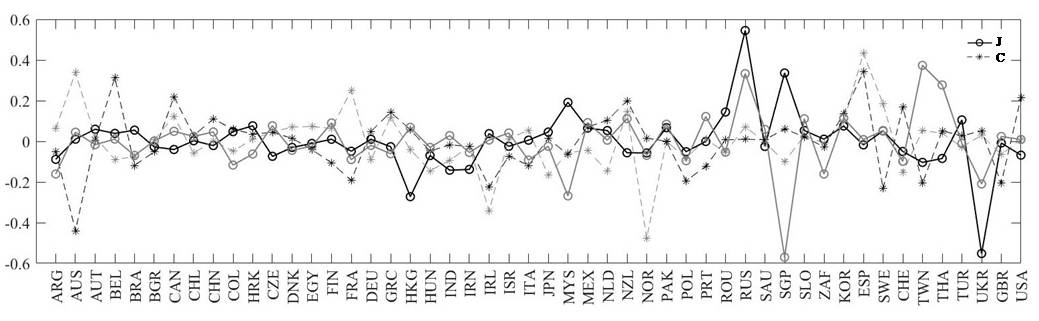}
\caption{Components of eigenvectors of $\textbf{J}$ (full line) and $\textbf{C}$ (dashed line) corresponding to the first two lowest eigenvalues.}\label{LowEig}
\end{figure*}
\begin{figure*}
\includegraphics[width=0.75\textwidth]{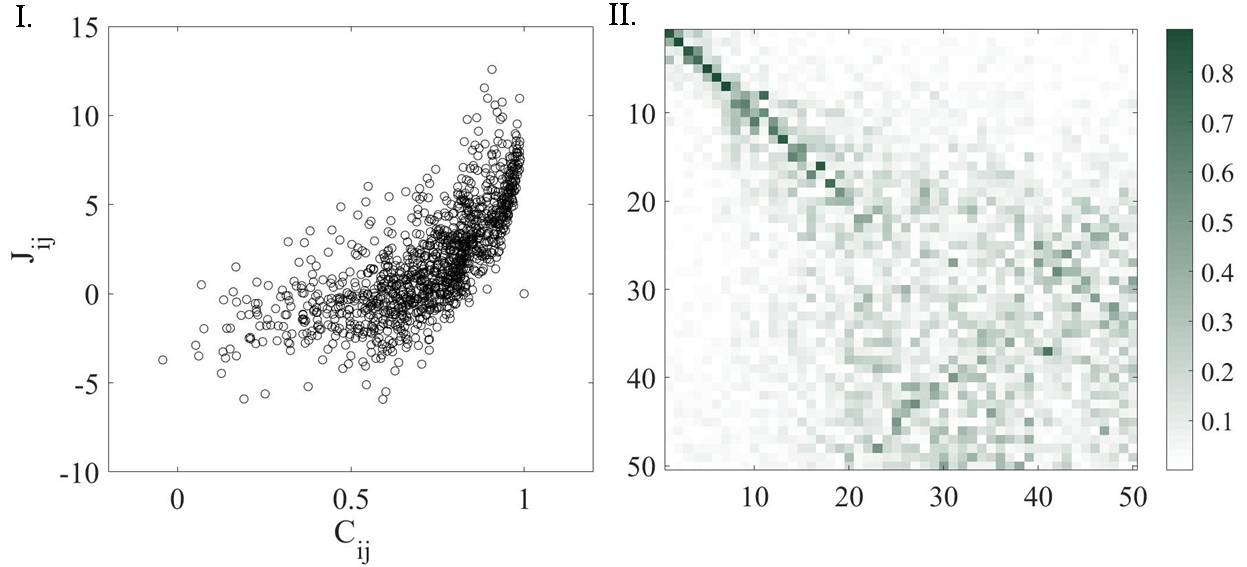}
\caption{\textit{Panel I.} Interactions $J_{ij}$ plotted against $C_{ij}$. \textit{Panel II.} Matrix reprsentation of the scalar product of the eigenvectors of $C$ and $J$, $|\bm{\xi}_i \cdot \bm{\zeta}_j|$ sorted by descending order of corresponding eigenvalue value.}\label{JeC}
\end{figure*}
The eigenvector corresponding to the largest eigenvalue is thus related to the whole structure of the interactions network, so in the case of empirical cross-correlations the eigenvector related to largest eigenvalue carry information about collective modes of the system. Similarly to the case of the empirical cross-correlations, the eigenvectors associated to the lowest eigenvalues are sensitive to the largest values of $J_{ij}$'s. Fig.\ref{LowEig} shows the eigenvector composition of $J$ (full lines) and $C$ (dashed lines) corresponding to the respective first two lowest eigenvalues. The first two largest components of the eigenvector related to the lowest $\lambda_J$ (full black line) correspond to Russia and Ukraine and the largest interaction $J_{ij}$ is the one between the same two countries. It exists a correlation between the values of $C$ and of $J$, as it is possible to infer by observing Fig.\ref{JeC}, \textit{Panels I}, which shows the values of $J_{ij}$ as a function of the values of $C_{ij}$. The nature of such a correlation can be better understood when the absolute value of the scalar product of the eigenvectors of the matrix $C$ and of the matrix $J$, i.e. $|\bm{\xi}_i \cdot \bm{\zeta}_j|$ is taken into account. Fig. \ref{JeC}, \textit{Panel II} shows the scalar product of couples of eigenvectors related to the two matrices ordered for increasing value of eigenvalues. The region corresponding to largest eigenvalues is near diagonal, outlining how for these eigenvalues the eigenvectors decomposition of $\textbf{C}$ and $\textbf{J}$ is similar. 
The eigenvectors of $\textbf{C}$ and $\textbf{J}$ related to lowest eigenvalues, which carry respectively information on the couples of countries strongly interacting and highly correlated, do not preserve such a correlation. 
This emphasizes that there is not a one-to-one correspondance between largest values of $J_{ij}$ and $C_{ij}$, see also Fig.\ref{LowEig}
\begin{figure*}
\includegraphics[width=0.99\textwidth]{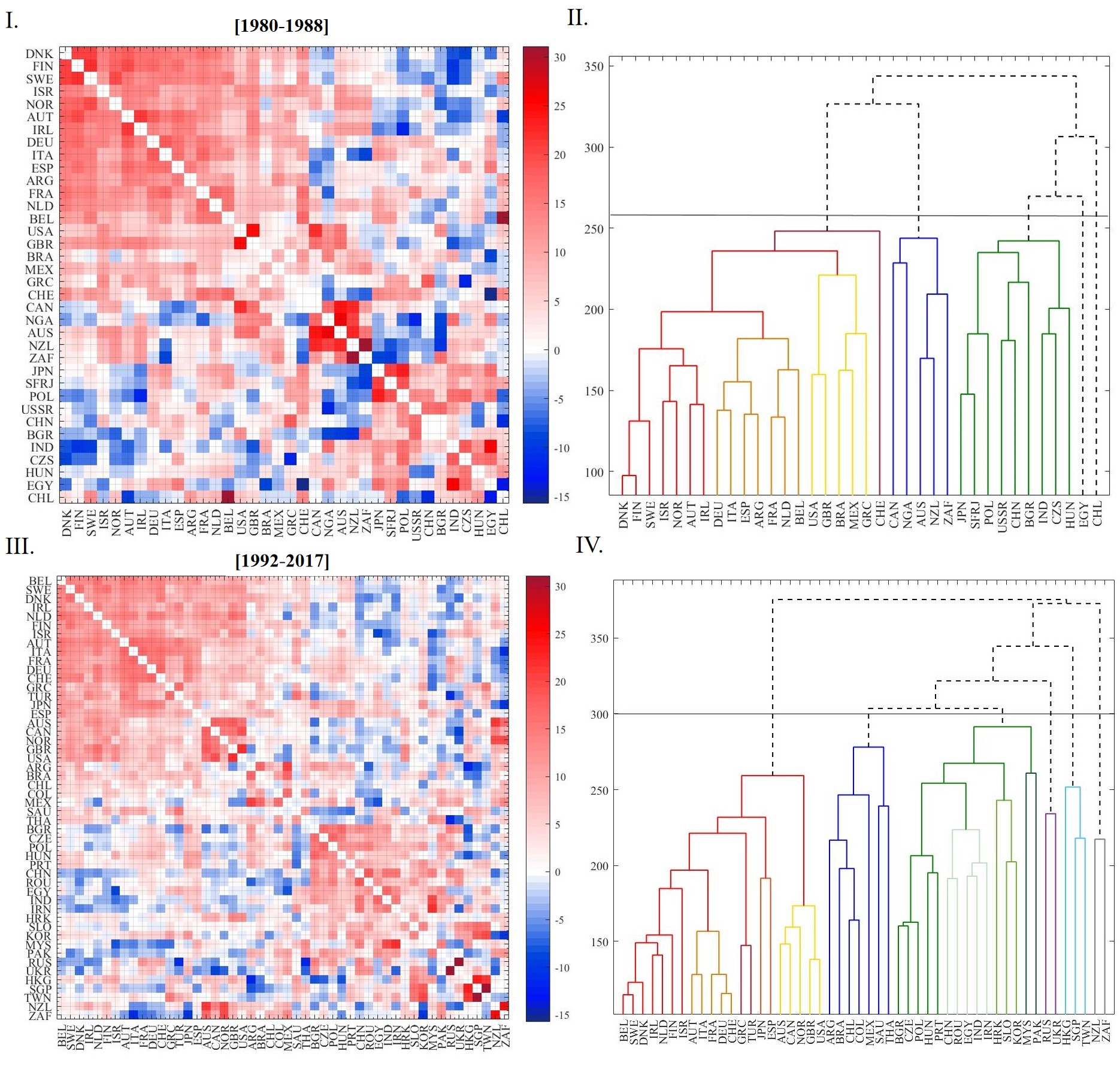}
\caption{\textit{Panel I.}Inferred interactions $J_{ij}$ for the period $[1980-1988]$. \textit{Panel II} Hierarchical clustering analysis of $\textbf{J}$ for the period $[1980-1988]$. \textit{Panel III.}Inferred interactions $J_{ij}$ for the period $[1992-2017]$. \textit{Panel IV.} Hierarchical clustering analysis of $\textbf{J}$ for the period $[1992-2017]$.}\label{HC_fig}
\end{figure*}
\begin{figure*}
\includegraphics[width=1\textwidth]{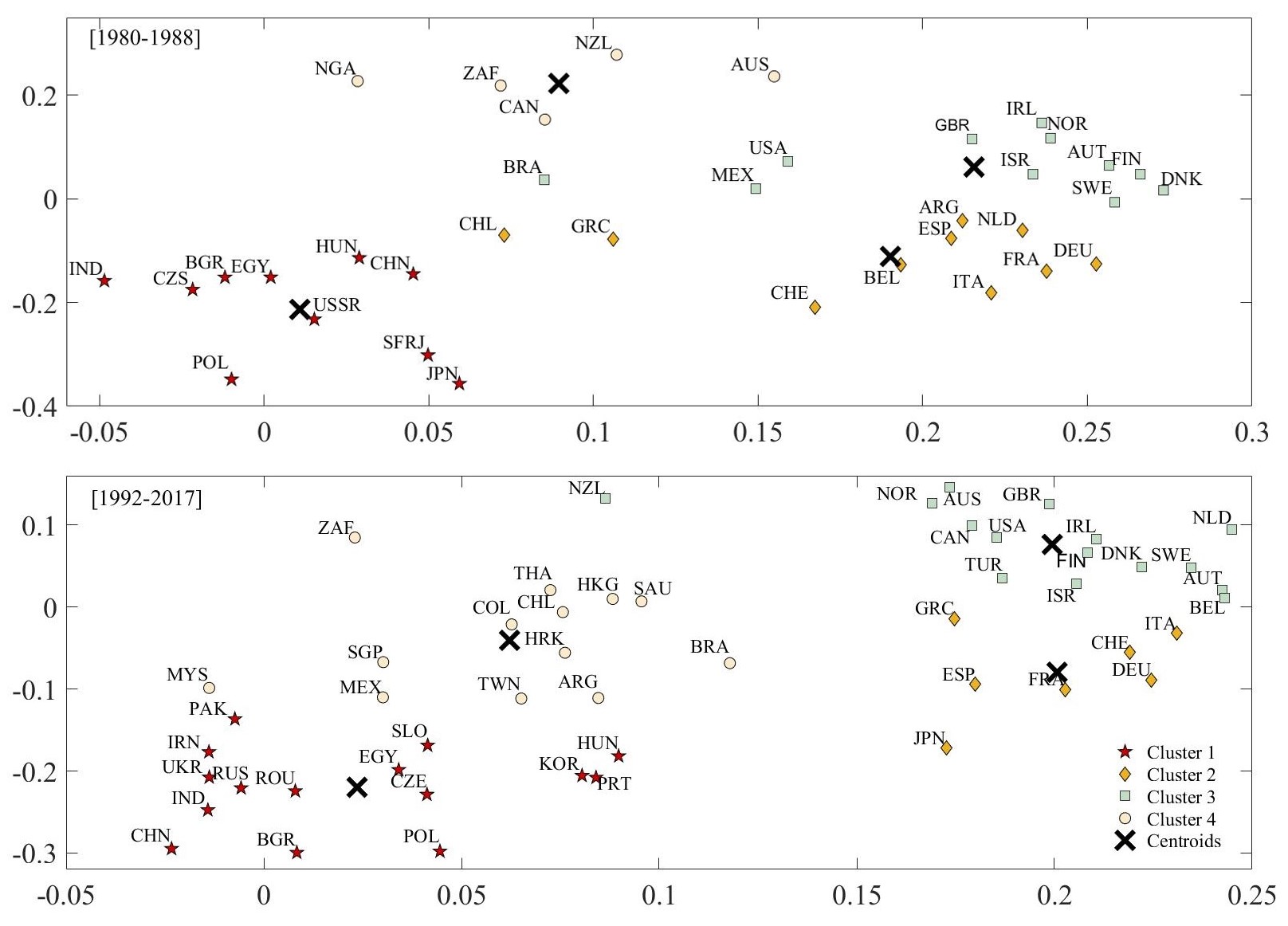}
\caption{PCA analysis of $\{ J\}$ in the time intervals $[1980-1988]$ and $[1992-2017]$. 2D graphycal representation of the eigenvectors related to the first two largest eingevalues of $\textbf{J}$. Clusters and clusters centroids are shown.}\label{PCApree88}
\end{figure*}
\begin{figure*}
\includegraphics[width=0.75\textwidth]{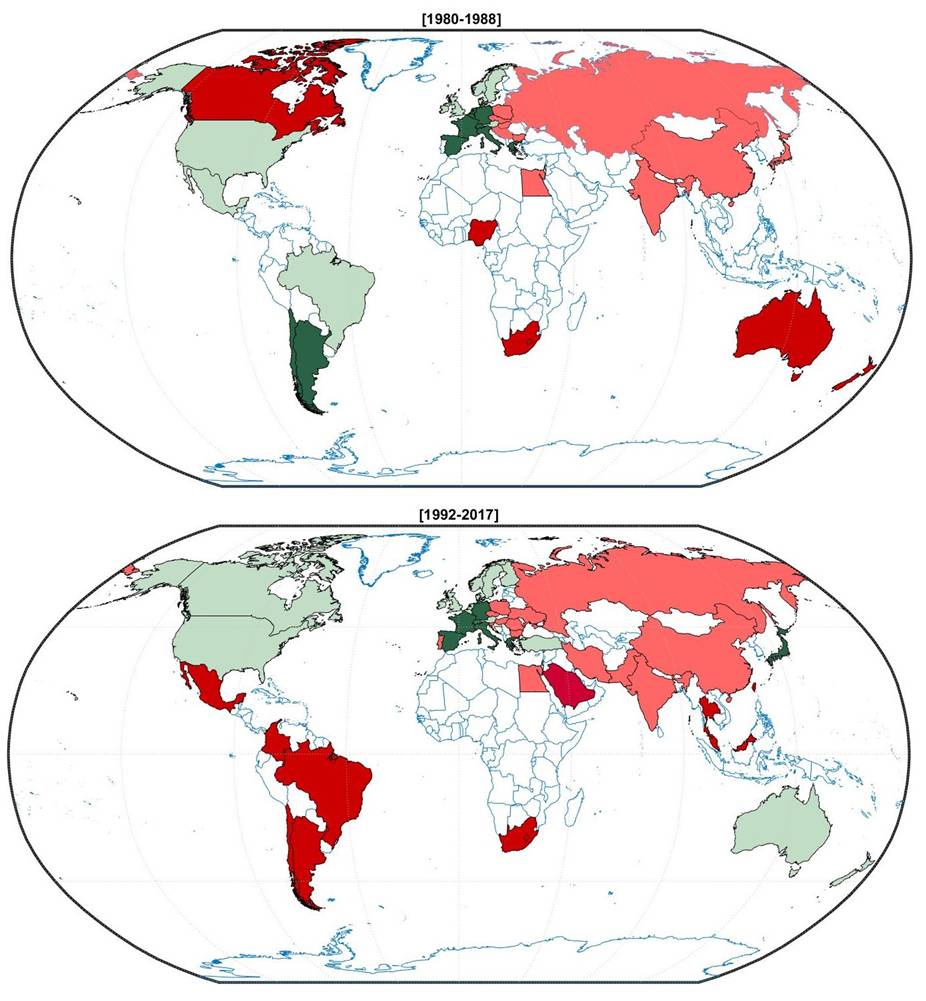}
\caption{Map representation of PCA clusters. }\label{PCAmaps}
\end{figure*}

\subsection{\textbf{Hierarchical clustering and principal components analysis of J}} \label{HC_PCA}
In the following we apply to the set of inferred interactions $\{ J \}$ two general methods usually applied to the analysis of correlation matrices, i.e. HC and PCA [\onlinecite{Barber1,Laio}]. 
Fig. \ref{HC_fig}, \textit{Panels} I/II show the elements of the inferred matrix $\textbf{J}$. For sake of clarity, the elements of the interaction matrix have been ordered following the HC outputs described in the following.

The HC is a hierarchical clusterization method [\onlinecite{Barucca}]. First, it is defined a metrics and it is calulated the distance between each two columns of the interaction matrix. In the present case the metrics adopted is city block distance, $d(\vec{J}_i,\vec{J}_j)=\sum_{\alpha}|J_{i\alpha}-J_{j\alpha}|$, where $\vec{J}_{i(j)}$ states for a column of $\textbf{J}$. At the starting step each column of $\textbf{J}$ corresponds to a different cluster. At each step the two clusters at the shortest distance are merged and form a cluster. This protocol allows to build a so-called dendrogram, the tree diagram shown in Fig. \ref{HC_fig}, \textit{Panels} II/IV. The height of the link between two objects, i.e. the countries displayed on the horizontal axis, indicates the distance between the objects. The dendogram can be cut at a given height thus allowing the definition of a certain number of clusters. 
The right panels of Fig. \ref{HC_fig} show the result of the HC protocol applied to $\textbf{J}$, different colors identify different clusters. 

The PCA is a partial eigendecomposition of the matrix $\textbf{J}$, where only the eigenvectors corresponding to the largest eigenvalues are considered [\onlinecite{Barber1,Laio}]. The selcted eigenmodes are those whose eigenvalues summed up descirbe with an error small enough the trace of the original matrix. Following this criterion we select the eigenvectors corresponding to larger (in magnitude) eigenvalues of $\textbf{J}$. As discussed in Sec. \ref{network} such eigenmodes bring genuine information, see also Fig. \ref{ProbJ}.   
A bidimensional representation of the eigenvectors of $\textbf{J}$ corresponding to the two largest eigenvalues is obtained by plotting in the bidimensional plane the points whose coordinates are the components of the two corresponding eigenvectors, see Fig. \ref{PCApree88}. This plot allows the identification of clusters of countries which have a similar interaction with the rest of the whole system [\onlinecite{Barucca}], as described in the following. The value of a given component of the eigenvectors related to the largest eigenvalues bring information on how the country associated to the given component interact with all the other countries, as also emphasised by the results on IPR described above. For example if the interaction network is such that only the interaction between countries $i$ and $j$ is different from zero, whereas all the others are zero, the eigenvector of the interaction matrix corresponding to the largest eigenvalue will have the only components $i$ and $j$ different from zero. A similarity criterion can be thus established: the two countries associated to the eigenvector's components which have a similar value (they are the only two different from zero) are those which have a similar among them, but different with respect to all the others, network of interactions (they are the only two interacting with at least one other country). For more complex cases, where the interaction matrix has more than one element different from zero, one can assume that those countries which are neighbors on the PCA plane are connected with other nodes with a similar interactions network. Notice that two countries similar following the PCA criterion can have a small interaction. A K-means clustering procedure with a squared euclidian metrics [\onlinecite{Barber1,Laio}] is used in order to obtain the clusters decomposition shown in Fig. \ref{PCApree88}. The number of the cluster is fixed to four. The centroids of the cluster and the attribution of a given point to a cluster are determined by minimizing the function $\chi_K=\sum_{k=1}^4\sum_{i \in C_k}d(x_i, c_k)$, where $i$ is the point's index, k indices the cluster $C_k$ and $d(x_i, c_i)$ is the distance between points and centroids in the specified metrics (euclidean in the present case). \textit{l2}-regularization have been introduced in the optimization routine (parameter 0.13). HC and K-means clustering have been performed by MATLAB packages [\onlinecite{Matlab}]. Fig. \ref{PCAmaps} finally shows in a geographical map the clusters obtained by PCA.

\section{\textbf{A geopolitical feedback}} \label{discusion}
Even if a geopolitical analysis of the results obtained (Figs. \ref{PCApree88} and \ref{PCAmaps}) is beyond the aim of the present work, we shortly point out in the following how the inferred results are in agreement with the general lines one can draw basing only on sinple geopolitical arguments concerning international relations in a global context both before and after the fall of the Berlin Wall. In the time interval $[1980-1988]$ the existence of a so-called Communistic block, including Soviet Union, Eastern Europe countries and China (cluster 1 in Fig. \ref{PCApree88}) can be clearly observed. Interestingly, India, China and Japan belong to this same cluster centered in the Soviet Union. The existence of a similar cluster is preserved in the period $[1992-2017]$, following the fall of the Berlin Wall, with the only relevant exception of Japan, which during the more recent time interval belongs to the Western block (cluster 2 in Fig. \ref{PCApree88}). Partnership between India, Soviet Union (or Russia) and China, either during the Cold War and after are well-documented [\onlinecite{Keylor, Kahan, Naik, Mastny, Singh}], also partially formalized in the so-called BRICS alliance [\onlinecite{Neill}]. The role of Japan to the Cold War and its position change possibly dating at its involvement to the Gulf war in 1991 [\onlinecite{Keylor}] and at the establishment of the Japanese-American Security Treaty in 1996 [\onlinecite{Keylor, Feske}] is also a well-assessed concept in the historical analysis.

The PCA analysis emphasizes the presence of other two clusters in the time period $[1980-1988]$: one including the Central and South Europe countries (cluster 2) and a second one including United Kindom, United States and Scandinavian countries (cluster 3). As it is possible to observe in Fig. \ref{PCApree88} the centroids of clusters 2 and 3 are closer and they are both distant from the centroid of cluster 1. This bipolar configuration is mostly preserved in the time period $[1992-2017]$, where the presence of one pole encompassing the North American and Europeans countries and the other one the former communist block countries, is still observable, despite the fall of the Berlin Wall. This result possibly supports the hypothesis that the conflict, which took shape in the Cold War and was also played on the plane of a territorial control, was anyhow mantained, after the fall of the Berlin Wall, on a socio-political, cultural and economic plane. It is also relevant to observe that in the post-Wall period the distance between United State and the Western Europe countries shortened. Furthermore, whereas in the period $[1980-1988]$ the Latin America countries belonged to the same cluster than United States and Western Europe countries, in the more recent time interval they aggregate in a separate cluster (cluster 4). This rearrangement is accompained by the rapprochement of United States to Western Europe countries, as noticed above, and, similarly, of Canada and Australia, which only in the time-period $[1992-2017]$ belong to the same Western countries cluster (cluster 3).

Finally in the following we analyse a bit more in details the European configuration observed in the two time period, in particular putting it in relation with the shaping of the European Community and NATO alliance. 
The single European Act of 1986 stipulated that by the beginning of 1993 free movement of goods, services, capital and labor among the twelve member states of the so-called European Community (EC) would be achieved. The EC refers to the association of countries from the European Economic Community, the European Coal and Steel Community and the European Atomic Energy Community taken place in 1967 [\onlinecite{Keylor}]. The original nucleus of EC are the six signatories of the Treaty of Rome (France, Germany, Italy, Belgium, the Netherlands, Luxembourg) plus Great Britain, Ireland, Denmark, Greece, Spain and Portugal. Accordingly, during the time interval $[1980-1988]$ the six original EC's countries all belong to cluster 2, together with Spain and Greece. Great Britain, Ireland and Denmark, although next-neighbors of the above-cited EC countries, belong to Cluster 1. Cluster 1 furthermore encompasses the states of the European Free Trade Association (EFTA), i.e. Austria, Finland, Norway and Sweden. Emerged in 1960 as a rival to the old European Economic Community, EFTA, apart from Norway, merged into the EC in 1995 [\onlinecite{Keylor}]. Negotiations with the four countries began in 1993.
Consistently with this scenario the core of the EC countries and the four EFTA countries belong to different, but next neighbors, clusters (cluster 2 and 3) in both the periods [1980-1988] and [1992-2017].
The members of NATO after 1992 are the twelve founding members (the United States, the United Kingdom, Belgium, Canada, Denmark, France, Iceland, Italy, Luxembourg, the Netherlands, Germany Norway, Portugal), Greece, Turkey, Spain, the former Warsaw Pact countries (the three Eastern European countries Hungary, the Czech Republic, Poland, ex members of COMECON, accessed in 1997, Bulgaria, Romania, Slovakia, Slovenia and the Baltic states Estonia, Latvia and Lithuania in 2004), Albania and Croatia entering in 2009 and Montenegro in 2017. As observed above, the inclusion of the Eastern European states in the NATO doesn't correspond to a migration of these states to the Western countries clusters in the time period $[1992-2017]$. This last point gives thought to the relations between NATO and Eastern Europe countries [\onlinecite{Paquette}].

This brief and essential overview, while not at all exhaustive, aims on one hand to highlight the reasonable agreement observed between the results obtained and an elementary geopolitical analysis and, on the other hand, to outline how this quantitative analysis can be a valuable instrument for a historical analysis.

\section{\textbf{Summary and Outlook}} \label{conclusion}
We defined the disciplinary profile of a country as a versor whose elements are the number of articles published by the given country in a given discipline divided the total number of articles published by the country. The countries considered are those supporting a significant number of published articles with respect to the worldwide production. Each country can be associated to the node of a graph. A partial definition of cultural production of a given country, i.e. only the one which take shape in articles production and recorded in the specified databases, is adopted. This has to be taken in mind if a historical analysis is done on the basis of the present results. Time series of country-level disciplinary profiles are acquired in the time intervals $[1980-1988]$ and $[1992-2017]$ with a time-step of a year. Since in between the two time intervals the fall of the Berlin Wall, with the related historical events, caused a reorganization of the geopolitical map, the graph under exam would need to be differently defined in the two time intervals. For each graph the set of the pairwise interactions has been inferred. A comparison between the results obtained in the two time intervals, however, can be of interest in a geopolitical perspective. A preliminary analysis of the empirical cross-correlation matrices of the disciplinary profiles in the more recent time interval has been performed. By exploiting a comparison with RMT results it was possible to establish that the cross-correlation matrices contain, beyond noise content, genuine information, identified in the largest and smallest eigenvalues and the corresponding eigenvectors. They are furthermore stationary in time. 
After proving that the empirical cross-correlation matrices bring genuine information, an inference procedure based on maximum entropy modeling of second-order marginal has been applied to the data in order to infer the value of pairwise interactions $J_{ij}$. The maximum entropy modeling is equivalent to the maximization of a Likelihood function belonging to the class of Boltzmann distribution related to a generalized Heisenberg model with $n_d$-dimensional spin variables. In the present case $n_d$ is equal to the number of disciplines considered. In order to obtain a working algorithm able to draw the optimal matrix of pairwise interactions we used a Pseudo-Likelihood maximization approach. We analytically computed the Pseudo-Likelihood and its gradient in order to facilitate the computational solution of the inference problem. The analytical computations reserve by themselves interest in Bayesian inference framework whenever a $n_d$-dimensional Heisenberg model is appropriate for the inference problem one aims to solve. 
We finally obtained the optimal value of the matrix $\textbf{J}$ by numerical maximization of the Log-Pseudo-Likelihood function. To the inferred interactions matrix they have been applied two classification methods, Hierarchical Clustering and Principal Component Analysis, usually exploited in the analysis of the cross-correlation matrices. We obtained a clusters representation of the interactions network shown in Figs. \ref{HC_fig}-\ref{PCAmaps}. 
An elementary geopolitical analysis of the results obtained emphasizes the soundness of the results obtained, it calls for deeper historical analysis and, finally, it support the use of physical modeling in this field.  

We finally point out two outlooks of the present work. Correlations among two variables $\textbf{s}_i$ and $\textbf{s}_j$ can be caused either by direct statistical coupling and indirect correlation effects, such it is the case, e.g., in the Heisenberg model of two variables $\textbf{s}_i$ and $\textbf{s}_j$ not interacting among themselves but both interact with a third variable $\textbf{s}_k$. The inference of the interactions matrix $\textbf{J}$ allows to identify variables statistically coupled. Once the matrix $\textbf{J}$ has been inferred it could be interesting to disentangle direct and indirect correlations.
Given two nodes $i$ and $j$ of a graph this can be achieved by defining the so-called direct information [\onlinecite{Morcos}] $DI_{ij}=\sum_{\{\textbf{s}_i,\textbf{s}_j\}}P^{(dir)}_{ij}(\textbf{s}_i,\textbf{s}_j)\log{\frac{P^{(dir)}_{ij}(\textbf{s}_i,\textbf{s}_j)}{f_i(\textbf{s}_i)f_j(\textbf{s}_j)}}$, where the sum is on the two-variables configuration space, $\{(\textbf{s}_i,\textbf{s}_j)\}$, and $f_{i}(\textbf{s}_i)=\sum_{\mu=1}^M\delta(\textbf{s}_i^{\mu}-\textbf{s}_i)$ is the frequency of observation in the time series $\{ \textbf{s}_i^{\mu}\}$ of the variable-value $\textbf{s}_i$. The so-called two-variable direct probability $P^{(dir)}_{ij}(\textbf{s}_i,\textbf{s}_j)$ is the key quantity in $DI_{ij}$. It can be obtained through the definition of so-called three messages $\nu_{i \rightarrow j}(\textbf{s}_i)$, $P^{(dir)}_{ij}(\textbf{s}_i,\textbf{s}_j)=\frac{\nu_{i \rightarrow j}(\textbf{s}_i)e^{-J_{ij}\textbf{s}_i \cdot \textbf{s}_j}\nu_{j \rightarrow i}(\textbf{s}_j)}{\sum_{\{ \textbf{s}_i,\textbf{s}_j\}}\nu_{i \rightarrow j}(\textbf{s}_i)e^{-J_{ij}\textbf{s}_i \cdot \textbf{s}_j}\nu_{j \rightarrow i}(\textbf{s}_j)}$ [\onlinecite{Morcos,Mezard}]. The message $\nu_{i \rightarrow j}(\textbf{s}_i)$ is the marginal distribution of the variable $\textbf{s}_i$ in a modified graph which does not include the node $i$. The message effectively represent the contribution to the correlation between the two variables not attributable to the direct coupling between them. Once $\textbf{J}$ is inferred, e.g. by exploiting the Pseudo-Likolihood approach as done in the present work, the three messages can be calculated through a self-consistent procedure, so-called belief propagation algorithm [\onlinecite{Mezard}]. 

Inference method based on maximization of the Likelihood function in order to infer the interactions $J_{ij}$ can be subject to overfitting due to the small number of collected configurations, i.e. limitness of the time-series. A way to avoid such a trouble is to use low-correlated data whose empirical cross-correlation function has an effective rank high enough [\onlinecite{Decelle2}]. Alternatively, the number of free parameters, i.e. $J_{ij}$ can be reduced, as done, e.g., in the so-called decimation algorithm [\onlinecite{Decelle3}]. In the mean field-approximation an expression of the Likelihood as a function of the eigenvalues of the correlation matrix is obtained [\onlinecite{Cocco}]. 
It is interesting to consider the explicit result obtained in Ref. [\onlinecite{Cocco}] for the Log-Likelihood function, $\log{l}=\sum_{i=1}^N\sum_{\{ \textbf{s}_i\}}f_i(\textbf{s}_i)\log{f_i(\textbf{s}_i)}+\frac{1}{2}\sum_n\lambda^{(n)}-1-\log{\lambda^{(n)}}$, where $\lambda^{(n)}$ is the $n$-th eigenvalue of $\textbf{C}$. From the expression above it follows that the single-eigenvalue contribution to the Log-Likelihood can be isolated and that the larger contribution comes from both largest and smallest eigenvalues of $\textbf{C}$. This is in agreement with the results of Sec. \ref{crosscorrelations} showing that these eigenvalues are those bringing the genuine information. It can be thus defined a hierarchy of eigenvalues according to their contribution to the Log-Likelihood function.
To each eigenvalue it is associated an eigenvector, whose non-null components define in the graph a pattern of nodes. Only pairwise interactions between the nodes belonging to the patterns associated to eigenvalues significantly contributing to the Log-Likelihood can be then the free parameters. This can give rise to a kind of piloted decimation.
The result sketched above, furthermore, emphasizes that in the mean-field approximation to the Log-Likelihood function they contribute not only the largest eigenvalues of $\textbf{C}$ but also the smallest. Performing PCA on the cross-correlation matrix allows to identify the path of nodes of maximum covariance. This is, however, not equivalent to find the path of maximally interacting nodes because, it is neglected the contribution of the lowest eigenvalues of $\textbf{C}$, which instead in the mean field-approximation contributes to the Log-Likelihood function [\onlinecite{Cocco}] and could anyhow affect the inferred $\textbf{J}$. This, together with the fact outlined above that correlations can be either direct or indirect, strengthens the need to go beyond analysis of empirical cross-correlation matrices and instead to infer $\textbf{J}$ if the aim one points to is to analyse the properties of the underlying interactions network generating the observed correlations. 

\section{Supplementary Note 1 - Eigenvalues of \textbf{C} and \textbf{C}(k)} \label{introduction}
\begin{figure}
\includegraphics[width=0.33\textwidth]{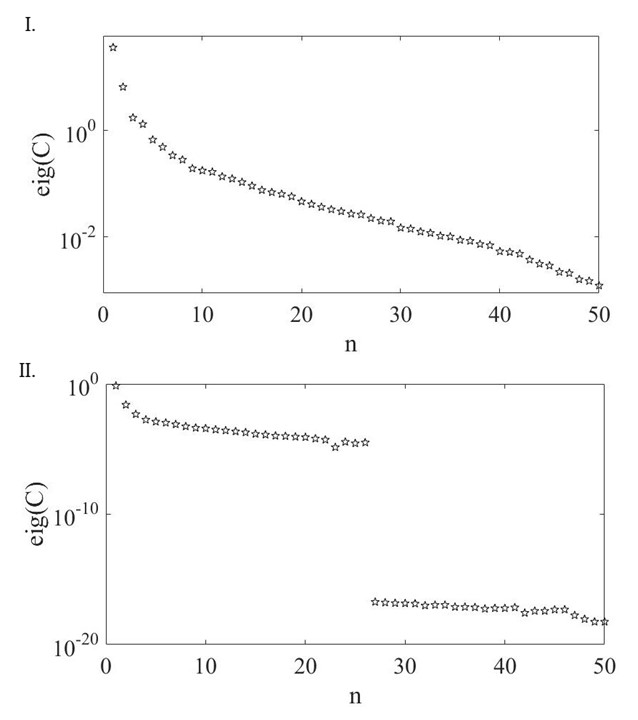}
\caption{\textit{Panel I}. Eigenvalues of the matrix $\textbf{C}$, sorted for ascending order. \textit{Panel II}. Eigenvalues of the single-discipline (MAT. SCI.) cross-correlation matrix $\textbf{C}(k)$ sorted for ascending order.}\label{eigenvalues}
\end{figure}

Fig. \ref{eigenvalues} emphasizes the difference observed between the case where $\textbf{s}_i$ is a scalar (single discipline) and a vector (disciplinary profile). In the scalar case the rank of the empirical cross-correlation matrix is $N-M$. Accordingly, the matrix $\textbf{C}(k)$ has $N-M$ null eigenvalues. This hampers the inference of couplings both in mean field approximation, where the couplings are obtained by inverting the empirical corss-correlation matrix, and in optimization procedure, e.g. minimization of the Chi-square function. In this latter case finite size of time series can lead to negative value of the averaged Chi-square when the number of free parameters is larger than the number of observed configurations. As shown in Fig. \ref{eigenvalues} in the vector case the matrix $\textbf{C}$ does not have null eigenvalues showing that, though correlations among different disciplines can occur, the number of degree of freedom of the time series of $\textbf{s}_i$ is increased. This would eventually restores a positive value of the averaged Chi-square and allows infrence procedure to be applied.

\begin{acknowledgments}
The authors thank A. Giansanti, M. Ibanez, V. Folli and G. Gosti for useful discussions.

\textbf{Authors Contribution}:
CD introduced to the research topic and provided the data. GR proposed the modeling. MGI did theoretical and numerical computations. MGI did data analysis and inference. MGI, CD, LL, GQ and GR discussed the results. MGI wrote the paper. CD and GQ wrote Sec. II.  MGI, CD, LL, GQ and GR revised the paper.   
\end{acknowledgments}

\end{document}